\documentclass{article}
\usepackage{jcappub}

\def\lbldef#1#2{\expandafter\gdef\csname #1\endcsname {#2}}

\def\href#1#2{#2}

\usepackage{graphicx}
\usepackage{epsfig}
\usepackage{amsmath}
\usepackage{amssymb}

\title{Improved constraints on the expansion rate of the Universe up to $z\sim1.1$ from the spectroscopic evolution of cosmic chronometers}

\author[1]{Moresco M.}
\author[1]{, Cimatti A.}
\author[2]{, Jimenez R.}
\author[3]{, Pozzetti L.}
\author[3]{, Zamorani G.}
\author[3]{, Bolzonella M.}
\author[4]{, Dunlop J.}
\author[5,6]{, Lamareille F.}
\author[3]{, Mignoli M.}
\author[4]{, Pearce H.}
\author[7]{, Rosati P.}
\author[8]{, Stern D.} 
\author[2]{, Verde L.}
\author[3]{, Zucca E.}
\author[9]{, Carollo C.~M.}
\author[5,6]{, Contini T.}
\author[10]{, Kneib J.-P.}
\author[10]{, Le~F\`evre O.}
\author[9]{, Lilly S.~J.}
\author[7]{, Mainieri V.}
\author[11]{, Renzini A.}
\author[12]{, Scodeggio M.}
\author[10,14]{, Balestra I.}
\author[13]{, Gobat R.}
\author[4]{, McLure R.}
\author[3]{, Bardelli S.}
\author[14]{, Bongiorno A.}
\author[4]{, Caputi K.}
\author[15]{, Cucciati O.}
\author[4]{, de~la~Torre S.}
\author[4]{, de~Ravel L.}
\author[12]{, Franzetti P.}
\author[12]{, Garilli B.}
\author[16]{, Iovino A.}
\author[9]{, Kampczyk P.}
\author[9]{, Knobel C.}
\author[17,7]{, Kova\v{c} K.}
\author[5,6]{, Le~Borgne J.-F.}
\author[10]{, Le~Brun V.}
\author[18,7]{, Maier C.}
\author[5,6]{, Pell\'o R.}
\author[9]{, Peng Y.}
\author[19]{, Perez-Montero E.}
\author[20,16]{, Presotto V.}
\author[21]{, Silverman J.~D.}
\author[21]{, Tanaka M.}
\author[10]{, Tasca L.~A.~M.}
\author[10]{, Tresse L.}
\author[22,3]{, Vergani D.}
\author[23]{, Almaini O.}
\author[9]{, Barnes L.}
\author[9]{, Bordoloi R.}
\author[23]{, Bradshaw E.}
\author[3]{, Cappi A.}
\author[23]{, Chuter R.}
\author[4]{, Cirasuolo M.}
\author[14,3]{, Coppa G.}
\author[9]{, Diener C.}
\author[24]{, Foucaud S.}
\author[23]{, Hartley W.}
\author[25]{, Kamionkowski M.}
\author[26]{, Koekemoer A.~M.}
\author[10]{, L\'opez-Sanjuan C.}
\author[27]{, McCracken H.~J.}
\author[3]{, Nair P.}
\author[9,28]{, Oesch P.}
\author[29,30]{, Stanford A.}
\author[31]{and Welikala N.}

\affiliation[1]{Dipartimento di Astronomia, Universit\'a di Bologna, via Ranzani 1, 40127 Bologna, Italy}
\affiliation[2]{ICREA \& ICC, University of Barcelona (IEEC-UB), Barcelona 08028, Spain}
\affiliation[3]{INAF - Osservatorio Astronomico di Bologna, via Ranzani 1, 40127 Bologna, Italy}
\affiliation[4]{SUPA, Institute for Astronomy, The University of Edinburgh, Royal Observatory Edinburgh, EH9 3HJ, United Kingdom}
\affiliation[5]{Institut de Recherche en Astrophysique et Plan\'etologie, CNRS, 14, avenue Edouard Belin, F-31400 Toulouse, France}
\affiliation[6]{IRAP, Universit\'e de Toulouse, UPS-OMP, Toulouse, France}
\affiliation[7]{European Southern Observatory, Karl-Schwarzschild-Stra{\ss}e 2, 85748 Garching bei M\"unchen, Germany}
\affiliation[8]{Jet Propulsion Laboratory, California Institute of Technology, Mail Stop 169-221, Pasadena CA-91109, USA}
\affiliation[9]{ETH Zurich, Institute of Astronomy, Wolfgang-Pauli-Stra{\ss}e 27, 8093 Zurich, Switzerland}
\affiliation[10]{Laboratoire d'Astrophysique de Marseille, Universit\'{e} d'Aix-Marseille, CNRS, 38 rue Fr\'{e}d\'{e}ric Joliot-Curie, 13388 Marseille Cedex 13, France}
\affiliation[11]{INAF - Osservatorio Astronomico di Padova, vicolo dell'Osservatorio 5, 35122 Padova, Italy}
\affiliation[12]{INAF - IASF Milano, via Bassini 15, 20133 Milano, Italy}
\affiliation[13]{Laboratoire AIM-Paris-Saclay, CEA/DSM-CNRS-Universit\'e Paris Diderot, Irfu/Service dÕAstrophysique, CEA Saclay, Orme des Merisiers, 91191 Gif-sur-Yvette, France}
\affiliation[14]{Max-Planck-Institut f\"ur Extraterrestrische Physik, Giessenbachstra{\ss}e, 85748 Garching bei M\"unchen, Germany}
\affiliation[15]{INAF - Osservatorio Astronomico di Trieste, via G. B. Tiepolo 11, 34143 Trieste, Italy}
\affiliation[16]{INAF - Osservatorio Astronomico di Brera, via Brera 28, 20121 Milano, Italy}
\affiliation[17]{Max-Planck-Institut f\"ur Astrophysik, Karl-Schwarzschild-Stra{\ss}e 1, 85748 Garching bei M\"unchen, Germany}
\affiliation[18]{University of Vienna, Department of Astronomy, Tuerkenschanzstra§e 17, 1180 Vienna, Austria}
\affiliation[19]{Instituto de Astrofisica de Andalucia, CSIC, Apdo. 3004, 18080 Granada, Spain}
\affiliation[20]{Universit\'a degli Studi dell'Insubria, Via Valleggio 11, 22100 Como, Italy}
\affiliation[21]{Institute for the Physics and Mathematics of the Universe (IPMU), University of Tokyo, Kashiwanoha 5-1-5, Kashiwa-shi, Chiba 277-8568, Japan}
\affiliation[22]{Istituto Nazionale di Astrofisica - Istituto di Astrofisica Spaziale e Fisica Cosmica Bologna, via P. Gobetti 101, I-40129 Bologna, Italy}
\affiliation[23]{University of Nottingham, School of Physics and Astronomy, Nottingham NG7 2RD}
\affiliation[24]{Department of Earth Sciences, National Taiwan Normal University, N¡88, Tingzhou Road, Sec. 4, Taipei 11677, Taiwan, Republic of China}
\affiliation[25]{Physics \& Astronomy Dept. John Hopkins University, 3400 N. Charles Street Baltimore, MD 21218, USA }
\affiliation[26]{Space Telescope Science Institute, 3700 San Martin Drive, Baltimore, MS 21218, USA}
\affiliation[27]{Institut d'Astrophysique de Paris, UMR 7095 CNRS, Universit\'e Pierre et Marie Curie, 98 bis Boulevard Arago, 75014 Paris, France}
\affiliation[28]{University of California Santa Cruz, UCO/Lick Observatory, 1156 High St, Santa Cruz, CA 95064, USA}
\affiliation[29]{Department of Physics, University of California, Davis, CA 95616, USA}
\affiliation[30]{Institute of Geophysics and Planetary Physics, Lawrence Livermore National Laboratory, Livermore, CA 94551, USA}
\affiliation[31]{Institut d'Astrophysique Spatiale, Batiment 121, CNRS \& Univ. Paris Sud XI, 91405 Orsay Cedex, France}

\emailAdd{michele.moresco@unibo.it}

\abstract{We present new improved constraints on the Hubble parameter $H(z)$ in the redshift range
$0.15<z<1.1$, obtained from the differential spectroscopic evolution of early-type galaxies as a function
of redshift. We extract a large sample of early-type galaxies ($\sim11000$) from several spectroscopic surveys, 
spanning almost 8 billion years of cosmic lookback time ($0.15<z<1.42$).
We select the most massive, red elliptical galaxies,
passively evolving and without signature of ongoing
star formation. Those galaxies can be used as standard cosmic chronometers, 
as firstly proposed by Jimenez \& Loeb (2002),
whose differential age evolution as a function of cosmic time directly probes $H(z)$.

We analyze the 4000~\AA~break ($D4000$) as a function of redshift,
use stellar population synthesis models to theoretically calibrate the dependence of the
differential age evolution on the differential $D4000$, and estimate the Hubble parameter
taking into account both statistical and systematical errors.

We provide 8 new measurements of $H(z)$ (see Tab. \ref{tab:HzBC03}),
and determine its change in $H(z)$ to a precision of $5-12\%$ mapping homogeneously the redshift 
range up to $z\sim1.1$; for the first time, we place a constraint on $H(z)$ at $z\neq0$ 
with a precision comparable with the one achieved for the Hubble constant (about 5-6\% at $z\sim0.2$),
and covered a redshift range ($0.5<z<0.8$) which is crucial to distinguish many different
quintessence cosmologies.

These measurements have been tested to best match a $\Lambda$CDM model, clearly 
providing a statistically robust indication that the Universe is undergoing an accelerated expansion.
This method shows the potentiality to open a new avenue in constrain a variety of alternative cosmologies,
especially when future surveys (e.g. Euclid) will open the possibility to extend it up to $z\sim2$.
}

\begin{document}

\maketitle

%%%%%%%%%%%%%%%%%%%%%%%%%%%%%%%%%%%%%%%%%%%%%%%%%%%%%%%%%%%%%%%%%%%%%%%%%%%%%%%%%%%%%%%%%%%

\section{Introduction}

The expansion rate of the Universe changes with time, 
initially slowing because of the mutual gravitational attraction of all the matter in it, 
and more recently accelerating, which is referred to generically as arising from 
``dark energy'' \cite{riess_de,perlmutter,dark1,dark2}.

The most generic metric describing a flat, homogeneous and isotropic Universe 
is the Friedmann-$\mathrm{Lema\hat{\i}tre}$-Robertson-Walker (FLRW) one: 
\begin{equation}
ds^2 = -c^2dt^2 + a(t) \delta_{i j} dx^i dx^j \nonumber
\end{equation}
that relates the line element in space-time ($ds^{2}$) to the time element ($c^2dt^2$) and to the 
space element ($dx^2$) using only the expansion factor $a(t)$, which characterizes
how space is expanding as a function of time. For a given model that specifies 
the equation of state of all components in the Universe, $a(t)$ is fully 
determined. 

However, we do not know what constitutes most of the energy budget in the Universe, 
and thus $a(t)$ needs to be determined observationally. The function $a(t)$
is related to the Hubble parameter by $H(t)=\dot a(t)/a(t)$. This
parameter has been measured with high accuracy ($\sim$3\%) only in 
the present-day Universe, i.e. the Hubble constant $H_0$ 
\cite{freedman,Riess,Moresco}. One of the key goals of modern cosmology is 
therefore to constrain 
$H$ as a function of cosmic time. To determine $H$, several observational tools
have been proposed, from standard ``candles'' (e.g. Type Ia 
Supernovae) to standard ``rulers'' (e.g. Baryonic Acoustic Oscillations), 
but none of them has achieved high accuracy results over a significant 
fraction of the Universe lifetime \cite{riesshz,daly,chuang}.

An independent approach is provided by the differential dating of {\it ``cosmic chronometers''}
firstly suggested by Jimenez \& Loeb (2002) \cite{JimenezLoeb2002}, 
because it gives a measurement of the expansion rate without relying
on the nature of the metric between the chronometer and us, which is not 
the case for methods which depend on integrated quantities along 
the line of sight. The cosmic chronometers formalism is very straightforward.

The expansion rate is defined as:
\begin{equation}
H(z) = \frac{\dot a}{a} = -\frac{1}{1+z} \frac{dz}{dt}
\label{eq:Hz}
\end{equation}
and since the redshift $z$ of the chronometers can be known with high accuracy (e.g. 
spectroscopic redshifts of galaxies have typical uncertainties $\sigma_z
\leq 0.001$), a differential measurement 
of time ($dt$) at a given redshift interval automatically provides a 
direct and clean measurement of $H(z)$. 

The major power of this method, as already underlined in Refs.~\cite{JimenezLoeb2002,Stern,Moresco},
is that it is based on a {\it differential approach}. This not only helps to cancel out 
the systematics that would have come in if evaluating absolute ages, but also
minimizes the potential effects of galaxy evolution: the integrated evolution as measured 
across all the redshift range it is not relevant when differential quantities are estimated, but all that matters is
just the evolution that takes place between the redshifts where the differences
are taken (for a more detailed discussion, see Sect.~\ref{sec:diff}).

If we want to move beyond the local Universe, the best cosmic chronometers
are galaxies which are evolving passively on a timescale much longer 
than their age difference. Based on a plethora of observational results, 
there is general agreement that these are typically massive (${\cal M}_{\rm stars} 
\sim 10^{11} {\cal M}_{\odot}$) early-type galaxies (ETGs hereafter) which formed the 
vast majority ($>$90\%) of their stellar mass at high-redshifts ($z >2-3$) 
very rapidly ($\sim$0.1-0.3 Gyr) and have experienced only minor subsequent 
episode of star formation, therefore being the oldest objects at all
redshifts (e.g. \cite{treu,renzini,thomas,pozzetti}). Thus, a 
differential dating of their stellar populations provides $dt$ in Eq.~\ref{eq:Hz}. 
It is worth recalling that differential dating of stellar populations 
is not only possible, but can be very accurate when targeting single stellar 
populations. As an example, we note that differential ages can be obtained for 
globular clusters in the Milky Way with a precision of 2-7\% (including 
systematic errors) (e.g. Ref.~\cite{gc}).

Compared to other approaches based on the global spectral or photometric analysis \cite{JimenezLoeb2002,Jimenez,simon,talavera,Stern,Capozziello2004}, 
it has been found that one of the most direct and solid ways of doing this is to use the 4000~\AA~break (hereafter $D4000$) in ETG 
spectra, thanks to its linear dependence on age for old stellar populations \cite{Moresco}. 
This break is a discontinuity of the spectral continuum around $\lambda_{\mathrm{rest}}$=
4000~\AA~due to metal absorption lines whose amplitude correlates 
linearly with the age and metal abundance (metallicity, $Z$) of the stellar 
population (in some age and metallicity ranges), that is weakly dependent (for old passive stellar populations) 
on star formation history ($SFH$), and basically not affected by dust
reddening \cite{Moresco,hamilton,Balogh1999,bc03} (see also Sect.~\ref{sec:model}, and figures therein).
If the metallicity $Z$ is known, it is then possible to measure the difference between the ages of two 
galaxies as proportional to the difference of their $D4000_n$ amplitudes:
$\Delta t = A(Z) \Delta D4000_n$, where $A(Z)$ is a slope which 
depends on metallicity. 

The differential aging of cosmic chronometers has been used to measure the observed
Hubble parameter \cite{Jimenez,simon}, to set constraints on the nature of dark energy 
\cite{Jimenez,simon,talavera}, and most 
recently to provide two new estimates of the Hubble parameter (even if with large errorbars) 
$H(z\sim0.5)=97\pm62\mathrm{\;km\;s^{-1}Mpc^{-1}}$ and 
$H(z\sim0.9)=90\pm40\mathrm{\;km\;s^{-1}Mpc^{-1}}$ \cite{Stern}, and
to recover the local Hubble constant \cite{Moresco}.

In this paper we present improved constraints on the Hubble
parameter up to redshift $z\sim1.1$, obtained using the technique described 
by Moresco et al. 2011 (\cite{Moresco}, hereafter M11). In order to fully exploit passive ETGs as reliable cosmic chronometers, two main
challenges must be faced: the appropriate sample selection and the reliable
differential dating of their stellar ages. 
The paper is organized as follows. The selection criteria and the
properties of the different samples are presented in Sect. \ref{sec:sample}.
In Sect. \ref{sec:D4000z} we introduce the theoretical basis used to estimate the Hubble 
parameter from the $D4000-z$ relation, describing how
the observed $D4000-z$ relation has been obtained and how stellar population
synthesis models have been used to calibrate the relation between $D4000$ and 
the age of a galaxy. In Sect. \ref{sec:Hzevaluation} we
discuss the detailed procedure to estimate $H(z)$, and how statistical and systematical
errors have been taken into account in the global error budget. In Sect.
\ref{sec:results} and \ref{sec:comparison} we present our $H(z)$ estimates, compare them with
other $H(z)$ measurements available in literature and show the constraints
our data impose on different cosmological scenarios.

\section{Sample selection}
\label{sec:sample}

For a reliable application of the {\it cosmic chronometers} approach, it is 
essential to select an appropriate sample of passively evolving 
ETGs over the widest possible redshift range.
The optimal choice to homogeneously 
trace the redshift evolution of cosmic chronometers would have 
been a dedicated survey, mapping with the same characteristics and 
properties the $D4000-z$ relation in the entire redshift range.
However, a single survey of ETGs covering 
a wide redshift range with spectroscopic information does not exist. 
To circumvent this limitation, we exploited both archival and still to be released surveys, and
the total sample used in this work is therefore the combination of several different subsamples. 

The general selection criteria adopted to extract the final sample of 
ETGs were based on the following main steps:

{\it (i)} extraction of the 
reddest galaxies with multi-band photometric spectral energy distributions 
(SEDs) compatible with the template SEDs of ETGs at $z\sim0$ or with old 
passive stellar populations \cite{Moresco2010};

{\it (ii)} high-quality optical spectra 
with reliable redshifts and suitable to provide D4000$_n$ 
amplitudes up to $z \sim 1.5$;

\begin{figure}[t!]
\begin{center}
\includegraphics[width=0.8\textwidth]{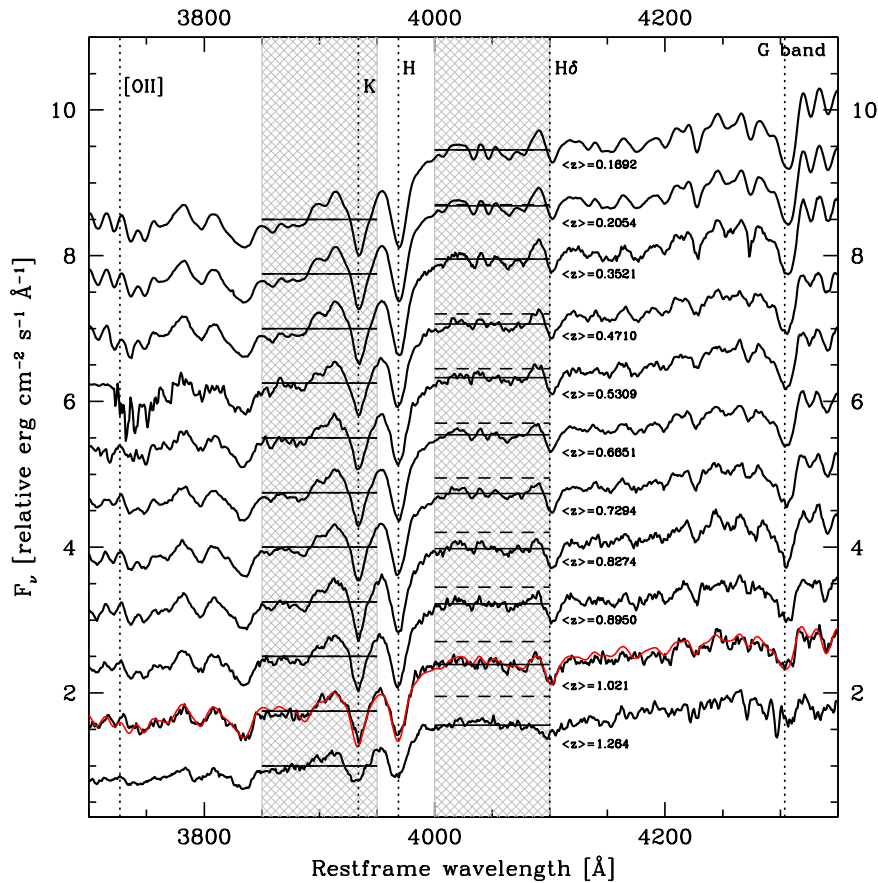}
\end{center}
\caption{The ETG spectral evolution. In order to
increase the signal-to-noise ratio and the visibility of spectral
features, mean stacked spectra were obtained by co-adding individual 
spectra of ETGs in each redshift bin. For each stacked spectrum,
the bin central redshift is indicated on the top-right. The
spectra are typical of passively evolving stellar populations and
do not show significant [O II]$\lambda$3727 emission. 
The spectra are normalized in the blue region of $D4000_{n}$ (3850-3950 {\AA}), where the average flux is 
indicated by a segment in the hatched region on the left. The hatched
region on the right indicates the red $D4000_{n}$ range (4000-4100 {\AA}), where
the solid segments represent the average fluxes and the dashed
one indicates the average flux of the lowest redshift spectrum.
A trend of decreasing red flux (i.e. $D4000_{n}$, which is defined as the ratio between the average fluxes in the 
red and blue ranges defined above) with increasing redshift
is clearly visible. As a reference, a BC03 spectrum with delayed $\tau$ SFH ($\tau=0.1$ Gyr), solar metallicity 
and age of 2.5 Gyr is overplotted in red to a high-z
stacked spectrum. The model spectrum has been convolved at a velocity dispersion
of 250 ${\rm km s^{-1}}$, typical of the ETGs considered.
\label{fig:stackspec}}
\end{figure}

{\it (iii)} absence of emission lines (H$\alpha$ and/or [OII]$\lambda$3727 
depending on the redshift) in order to exclude ongoing star formation or AGN activity; 
it is worth noting that emission lines (and in particular the [OII] and H$\alpha$ lines) are not detectable even if 
we average (stack) together the spectra of different ETGs in order 
to increase the signal-to-noise ratio (see Fig.~\ref{fig:stackspec}), hence excluding the 
possibility of low-level star formation or AGN activity not detected in 
individual spectra because of the higher noise. 

{\it (iv)} stellar masses ($\cal M$) estimated from photometric SED
fitting to be above $10^{11} {\cal M}_{\odot}$ (above $10^{10.6} {\cal M}_{\odot}$ at $z>0.4$) 
in order to select the most massive ETGs;

{\it (v)} spheroidal morphology typical of elliptical 
galaxies (when this information was available).\\

There now exists overwhelming evidence confirming a ``downsizing scenario'' for ETGs, 
with more massive ETGs having completed their star formation and mass assembly at higher
redshifts than less massive ones (e.g. see \cite{Fontana2004,treu,renzini,refdown1,refdown2,thomas,pozzetti,Moresco2010}). 
With the described selection criteria, thus, we have considered the reddest, oldest, passive envelope of ETGs in the entire 
redshift range, i.e. the best possible to trace the differential age evolution of the Universe.

Stellar masses were all evaluated assuming a standard cosmology, and rescaled (when necessary) to a Chabrier initial
mass function (IMF) \cite{Chabrier2003}. Given the non-uniform photometric and spectral coverage of the various surveys,
a totally homogenous mass estimate was not obtainable, and different models and techniques have been used.
However, this fact does not pose a major concern for the analysis, since different techniques recover very similar stellar masses 
for passively evolving galaxies (e.g. see \cite{pozzetti,Bolzonella2010}), and the primary parameter which may significantly bias the estimate 
is the IMF, which has been corrected for. Moreover it is worth emphasize that the masses estimated do not directly affect the scientific results, but are only
used to select the most massive galaxies in all the surveys, independently on their absolute value. Three high redshift early-type
galaxies (with $1.8<z<2.2$) have been also considered, studying the possibility to extend this approach up to much higher redshifts.

Each spectroscopic survey used for this purpose has its own characteristics. 
In the following, the relevant details of each subsample are presented
(see also Table \ref{tab:summarysurveys}).

\paragraph{{\bf SDSS-DR6 MG Sample.}} This sample has been taken from the analysis 
of M11. ETGs have been extracted from the 
SDSS-DR6 Main Galaxy Sample (MGS, \cite{Strauss2002}), matching SDSS photometry ($u$, $g$, $r$, $i$, and $z$) to 2MASS 
photometry ($J$, $H$, and $K$), in order to obtain a wider photometric 
coverage and extract robust mass estimates from the fitting of their 
photometric SEDs. For each galaxy, the 4000 ~\AA~ break amplitudes have been 
taken from the MPA-JHU DR7 release of spectral measurements\footnote{http://www.mpa-garching.mpg.de/SDSS/DR7/}. 
Passive ETGs have been 
selected combining a photometric criterion, i.e. selecting those galaxies 
whose best-fit to the SED matched a local E/S0 template, and a spectroscopic 
criterion by excluding those galaxies showing emission lines (rest-frame 
equivalent width EW$>5$~\AA). The stellar masses of these galaxies 
have been estimated with SED fitting, using a wide library built with 
BC03 models \cite{bc03}, exponentially delayed Star 
Formation Histories (hereafter SFH), with a Star Formation Rate (hereafter 
SFR) $SFR(t)\propto t/\tau^{2}\,exp(-t/\tau)$ with $0.05<\tau<1$ Gyr, 
ages with $0<t<20$ Gyr, dust reddening $0<A_{V}<1$ modeled with a 
Calzetti's extinction law \cite{Calzetti2000} ($0<A_{V}<0.6$ in the case of values of age$/\tau>4$), 
solar metallicity, and a Chabrier IMF. 
Despite the wide range of extinction allowed, the best-fit provided a distribution
of $A_{V}$ peaked at 0, with a median value of 0.2, compatible with the
selection of passive ETGs.
A mass cut has been applied, selecting galaxies with stellar masses 
$11<log({\cal M/M}_{\odot})<11.5$. 
Stellar metallicities were obtained from the estimates of Ref.~\cite{Gallazzi2005}\footnote{ 
www.mpa-garching.mpg.de/SDSS/DR4/Data/stellarmet.html}, who 
performed simultaneous fits of five spectral absorption features which 
depend negligibly on the $\alpha$/Fe ratio, i.e. $D4000$, 
H$\beta$ and H$\delta_{a}$+H$\gamma_{a}$ as age-sensitive indices and 
[Mg$_{2}$Fe] and [MgFe]' as metallicity-sensitive indices; some works (e.g. see \cite{Thomas2004,Korn2005}) 
have found in particular for H$\delta_{a}$+H$\gamma_{a}$ a dependence on $\alpha$/Fe ratio, but
in Ref. \cite{Gallazzi2005} it is also shown that the metallicities and ages obtained including or excluding those features do not present any
discrepancy. The original redshift range 
($0.15<z<0.3$, see M11 for further details) has been reduced to $z<0.23$ 
to limit the effect of the mass incompleteness due to the magnitude limit of the sample. In conclusion, the 
SDSS MGS ETGs sample contains 7943 ETGs in the redshift range $0.15<z<0.23$.

\paragraph{{\bf SDSS-DR7 LRGs sample.}} The Luminous Red Galaxies (LRGs) \cite{Eisenstein}
represent a spectroscopic sample of galaxies based on color and magnitude selection criteria, defined 
to yield a sample of luminous intrinsically red galaxies that extends fainter 
and farther than the main flux-limited portion of the SDSS main galaxy spectroscopic sample \cite{Abazajian2009}.
They are selected by imposing a luminosity and rest-frame color cut intended to follow passive evolution.
Two different cuts have been designed to select LRGs at $z\gtrsim0.4$ and $z\lesssim0.4$ 
(for further details, see \cite{Eisenstein}). 
Due to the informations available for this sample, it was not possible to apply the same spectro-photometric selection 
criterion used for the SDSS MGS; therefore, to reduce the possible contamination of starforming galaxies, we impose a threshold in the
signal-to-noise per pixel ratio for the spectra of these galaxies, rejecting galaxies with $S/N<3$,
and more severe spectroscopic cuts excluding galaxies with measured equivalent widths of the emission lines [OII] 
and H$\alpha$; as for the SDSS MG sample, the estimates of the equivalent widths and the S/N ratio have been 
taken from the MPA-JHU DR7 release of spectral measurements\footnote{http://www.mpa-garching.mpg.de/SDSS/DR7/}.
Stellar mass measurements of this sample have been obtained from VESPA \cite{Tojeiro2009},
a code developed to recover robust estimates of masses and star formation histories from 
a fit to the full spectral range of a galaxy with theoretical models; BC03 models have been adopted, 
and a Chabrier IMF. Due to their targeting criteria, LRGs sample a range of stellar masses higher 
than the previous SDSS ETGs, with $11<log({\cal M/M}_{\odot})<13$.
Taking into account both the effect of the mass incompleteness and 
the mass distribution of the sample, we decided to consider only the LRGs sample 
at $0.3<z<0.4$, and we selected galaxies with stellar masses $11.65<log({\cal M/M}_{\odot})<12.15$. 
In this way we obtained 2459 ETGs, in the redshift range $0.3<z<0.4$.

\paragraph{{\bf Stern et al. (2010) sample.}} This sample has been obtained 
from the analysis of Ref.~\cite{Stern}. Within this work, optical spectra 
of bright cluster elliptical galaxies have been obtained with the Keck 
LRIS instrument. Rich galaxy clusters were targeted in order to obtain 
an as large as possible sample of red ETGs over the redshift range 
$0.2<z<1$. Nine high S/N stacked spectra in the redshift range $0.38<z<0.75$ have been 
selected and analyzed (see Fig. 7 of Ref.~\cite{Stern}) to study the 
$D4000_{n}-z$ relation. All of these spectra clearly show features and 
continuum characteristic of old passive stellar populations.

\paragraph{{\bf zCOSMOS 20k bright sample.}} This sample has been extracted 
from the zCOSMOS 20k bright sample \cite{refzco}. The observed magnitudes 
in 12 photometric bands (CFHT $u^*$, $K$ and $H$, Subaru $B_J$, $V_J$, $g^+$, 
$r^+$, $i^+$, and $z^+$, UKIRT $J$ and Spitzer IRAC at 3.6 $\mu$m and 
4.5 $\mu$m) have been used in order to derive reliable estimates of galaxy
parameters from the photometric SED-fitting. The spectra have been obtained using the 
VIMOS spectrograph mounted at the Melipal Unit Telescope of the VLT at ESO's 
Cerro Paranal Observatory. The 4000~\AA~ break amplitudes 
have been obtained using the spectral measurements of Platefit \cite{Lamareille2006}.
Passive ETGs have been selected by combining photometric, 
morphological and optical spectroscopic criteria, following the approach 
of Ref.~\cite{Moresco2010}. Galaxies have been chosen with a reliable redshift 
measurement, a best-fit to the SED matching a local E-S0 template, weak/no 
emission lines (EW$<5$~\AA), spheroidal morphology, and a $K-24\mu$m color 
typical of E/S0 local galaxies (i.e. $K-24\mu$m$<-0.5$); for 
further details about the sample selection, see Ref.~\cite{Moresco2010}. 
The stellar mass has been estimated from SED fitting of those galaxies, 
using a wide library built with BC03 models, exponentially delayed SFHs with
$SFR(t)\propto t/\tau^{2}\,exp(-t/\tau)$ with $0.05<\tau<1$ Gyr, 
ages with $0<t<20$ Gyr, dust reddening $0<A_{V}<1$ modeled with a 
Calzetti's extinction law ($0<A_{V}<0.6$ in the case of values of age$/\tau>4$), solar metallicity, and a Chabrier IMF. 
As in the SDSS MGS sample, also in this case the best-fits to the data
presented a distribution of $A_{V}$ peaked at 0, with a median value of 0.2, 
compatible with the selection of passive ETGs. A mass cut 
$log({\cal M/M}_{\odot})>10.6$ has been applied to select the most massive 
population. Because of the wavelength coverage of the zCOSMOS spectra, 
the $D4000_{n}$ break is available only in the range $0.43\lesssim z\lesssim1.2$.
In conclusion, the zCOSMOS 20k ETGs sample contains 746 ETGs in the 
redshift range $0.43<z<1.2$.

\paragraph{{\bf K20 sample.}} The starting sample consists of about 500 
galaxies selected in the $K$-band from a sub-area of the Chandra Deep 
Field South (CDFS)/ GOODS-South and from a field around the quasar 
0055-2659 \cite{Cimatti2002}. Optical spectra were obtained with the ESO 
VLT UT1 and UT2 equipped respectively with FORS1 and FORS2.
Passive ETGs have been selected using 
the optical spectroscopic classification of Ref.~\cite{mignoli}, using a 
parameter \textsf{cls}=1, characteristic of red galaxies with no
emission lines and elliptical morphology. Mass estimates have been 
taken from the SED fitting of Ref.~\cite{Fontana2004}, who used a wide 
library of BC03 models (with exponentially decaying SFHs, $SFR(t)\propto 
1/\tau\, exp(-t/\tau)$, with $0.1<\tau<15$ Gyr, ages in the range 
$10^{7}<t<10^{10.2}$ yr, dust reddening $0<E_{B-V}<1$ modeled with 
SMC law, metallicities in the range $0.02<Z/Z_{\odot}<2.5$, and a 
Salpeter IMF). The stellar masses were rescaled to a Chabrier IMF 
by subtracting 0.23 dex from $log \cal M$ (see \cite{pozzetti,Bolzonella2010}), 
and selected to have $log({\cal M/M}_{\odot})>10.6$.
In conclusion, the K20 ETGs sample contains 50 galaxies 
in the redshift range $0.26<z<1.16$.

\paragraph{{\bf GOODS-S sample.}} Old passive ETGs were extracted from 
the GOODS-S field \cite{vanzella} combining morphological and 
photometric criteria based on optical color cuts as a function of redshift 
(for more details on the sample selection, see Ref.~\cite{Balestra2011}).
The spectra have been obtained from VVDS \cite{LeFevre2005}, VIMOS \cite{Balestra2010}, and
FORS2 \cite{vanzella,Szokoly2004}. Mass estimates have been taken from 
Ref.~\cite{Santini2007}, where the adopted Salpeter IMF was rescaled to a Chabrier 
IMF as previously discussed. A mass cut of $log({\cal M/M}_{\odot})>10.6$ 
has been used, and all galaxies with emission lines have been excluded 
as in the other samples. With this approach, 46 galaxies were selected
in the redshift range $0.67<z<1.35$.

\paragraph{{\bf Cluster BCGs sample.}} This sample consists of 
ETGs of the X-ray selected clusters RX J0152.7-1357 at $z=0.83$ \cite{Demarco2010}, 
RDCS J1252.9-2927 at $z=1.24$ \cite{Holden2005}, and XMMU J2235.3-2557 
at $z=1.39$ \cite{Rosati2009}, which include their BCGs and other galaxies within 250 kpc 
radius from the center, with no detectable [OII]$\lambda3727$ emission line in their spectra. 
Stellar masses have been evaluated from the SED fitting assuming BC03 models, solar 
metallicity, delayed exponential SFHs, and a Salpeter IMF; the masses have been therefore
rescaled to a Chabrier IMF as previously discussed.
By selecting spectra with high signal-to-noise, we are left with 5 galaxies in the range 
$0.83<z<1.24$, all with masses $log({\cal M/M}_{\odot})>11$.

\paragraph{{\bf GDDS sample.}} Within the GDDS \cite{Abraham2004}, 
Ref.~\cite{refgdds} analyzed the spectra (obtained with the GMOS multi-slit spectrograph) of 25 galaxies with measured 
$D4000_{n}$ and $H\delta$, in the range $0.6<z<1.2$ and with masses ${\cal
M}>10^{10.2}{\cal M}_{\odot}$. Stellar masses have been derived from template fits 
to the multicolor photometry \cite{Glazebrook2004}, assuming a 
Baldry et al. (2003) IMF \cite{Baldry2003}. Masses have been scaled by
-0.03 dex to convert them to a Chabrier IMF 
\cite{refgdds,Bolzonella2010}. ETGs have been selected to have a negligible 
specific Star Formation Rate ($sSFR=SFR/{\cal M}$, $sSFR<10^{-1}$
Gyr$^{-1}$), additionally applying a mass cut $log({\cal M/M}_{\odot})>10.6$. 
In this way, 16 galaxies were selected in the redshift range $0.91<z<1.13$.

\paragraph{{\bf UDS sample.}} Based on selection from the UKIDSS Ultra Deep Survey (UDS, \cite{refuds}), 
a spectroscopic survey was undertaken using the VIMOS and FORS2 spectrographs at the VLT 
(UDSz; \cite{Almaini2012}). From a spectroscopic sample of over 2000 galaxies, passive ETGs 
were identified using the criterion described in Ref.~\cite{Fontana2009}, with SFRs 
estimated using the rest-frame UV flux / [OII]$\lambda3727$ emission / 
24 $\mu$m detections. The stellar mass estimates were made by fitting 
double-burst models. Using Charlot \& Bruzual (2007), models were constructed as two sequential bursts, giving the galaxy 
an older and younger stellar subpopulation, assuming a Chabrier IMF. 
The models used for this analysis are the most dissimilar with respect to the
ones used in the other surveys; however, recently Ref. \cite{McLure2012}
estimated the stellar masses of a sample of massive galaxies in the UDS survey
with the same models, and comparing them with masses obtained with standard
$\tau$ BC03 model found a mean difference of only $\sim0.04$ dex (see Fig. 3 of Ref. \cite{McLure2012}).
The ETGs have been selected applying a mass cut $log({\cal 
M/M}_{\odot})>10.6$. A redshift cut $z<1.4$ has been applied, since 
at $z\geq1.4$ the $D4000_{n}$ break is near the red edge of the 
optical spectra, $\sim1\mu$m, where the sky noise is too high and CCDs become
transparent. In conclusion, this UDS sample contains 50 galaxies in the redshift 
range $1<z<1.4$.

\paragraph{{\bf High-z sample.}} We decided to expand our sample of ETGs
with three high redshift ($z>1.8$) galaxies. Onodera et al. (2010) \cite{Onodera2010} provided observations 
of a very massive galaxy at $z=1.823$ that shows properties which are fully consistent 
with those expected for passively evolving progenitors of today's giant ellipticals.
This high redshift galaxy has a stellar mass $log({\cal M/M}_{\odot})=11.37$. 
Ferreras et al. (2011) \cite{Ferreras2011} studied the properties of a massive (${\cal M/M}_{\odot}=0.8-3\cdot10^{11}$)``red and dead'' galaxy at 
$z=1.893$. Kriek et al. (2009) \cite{Kriek2009} have obtained a high-quality spectrum of a quiescent, 
ultra-dense galaxy at $z=2.1865$, with a stellar mass of $log({\cal M/M}_{\odot})\sim11.4$.
With this sample, we decided to test if the method can be extended up to much higher redshifts.\\

\begin{table}[h!]
\begin{center}
\begin{tabular}{lllll}
\hline \hline
survey & \# of galaxies & redshift range & mass range & Ref.\\
\hline
SDSS-DR6 MGS & 7943 & $0.15-0.23$ & $10^{11}-10^{11.5}{\cal M}_{\odot}$ & \cite{Moresco}\\
SDSS-DR7 LRGs & 2459 & $0.3-0.4$ & $10^{11.65}-10^{12.15}{\cal M}_{\odot}$ & \cite{Eisenstein}\\
Stern et al. sample & 9* & $0.38-0.75$ & - & \cite{Stern}\\
zCOSMOS 20k & 746 & $0.43-1.2$ & $10^{10.6}-10^{11.8}{\cal M}_{\odot}$ & \cite{refzco}\\
K20 & 50 & $0.26-1.16$ & $10^{10.6}-10^{11.8}{\cal M}_{\odot}$ & \cite{mignoli}\\
GOODS-S & 46 & $0.67-1.35$ & $10^{10.6}-10^{11.5}{\cal M}_{\odot}$ & \cite{vanzella}\\
Cluster BCG & 5 & $0.83-1.24$ & $10^{11}-10^{11.3}$ & \cite{Demarco2010,Holden2005,Rosati2009}\\
GDDS & 16 & $0.91-1.13$ & $10^{10.6}-10^{11.3}{\cal M}_{\odot}$ & \cite{refgdds}\\
UDS & 50 & $1.02-1.33$ & $10^{10.6}-10^{11.6}{\cal M}_{\odot}$ & \cite{refuds}\\
High-z sample & 3 & $1.8-2.2$ & $10^{11}-10^{11.4}{\cal M}_{\odot}$ & \cite{Onodera2010, Ferreras2011,Kriek2009}\\
\hline \hline
\end{tabular}
\caption{Summary table of the selected samples. *Stacked spectra of galaxies in 9 different rich galaxy clusters.}
\label{tab:summarysurveys}
\end{center}
\end{table}
 
With this approach, we selected a final sample of 11324 old, passively 
evolving ETGs with no signatures of star formation or AGN activity at $0.15<z<1.42$, and 
stellar masses in the range of $10.6<log({\cal M/M}_{\odot})<12.15$.
The key feature of our sample is the combination of the selection of {\it massive} and {\it passive} 
galaxies. This sample selection has been chosen to provide 
ETGs that formed most of their stars at comparable epochs (i.e. with similar redshift of formation \cite{thomas}), 
which means that they represent an homogeneous sample in terms of ages of formation, and can be used
as ``cosmic chornometers''. Moreover, basing on the 
stellar mass function \cite{pozzetti} and clustering \cite{masjedi08}, it has been found that this population of galaxies 
experienced negligible major merging events in a large fraction of the redshift range 
considered in our study \cite{nipoti}, i.e. they did not increase their mass 
significantly in the time comprised between the redshifts adopted for the differential approach,
meaning that the contamination to the initial population (with a given age of formation) is minimal.

We decided to treat separately the SDSS MG sample, the LRGs sample, 
and the sample at $z>0.4$, where, due to the lower statistical power 
provided by the existing surveys, we chose to merge together all of the high-redshift samples.
In Ref.~\cite{Moresco} it has been shown that there is an evident dependence of the $D4000$ break on the
stellar mass; therefore, to avoid spurious mass-dependent effects, we divided each subsample into
two mass bins, as follows:
\begin{itemize}
\item for the SDSS MG sample, we defined a low mass range for $11<log({\cal M/M}_{\odot})<11.25$ (5210 galaxies)
and a high mass range for $11.25<log({\cal M/M}_{\odot})<11.5$ (2733 galaxies, using the same mass ranges adopted in Ref.~\cite{Moresco});
\item for the LRGs sample, we defined a low mass range for $11.65<log({\cal M/M}_{\odot})<11.9$ (1410 galaxies) 
and a high mass range for $11.9<log({\cal M/M}_{\odot})<12.15$ (1049 galaxies);
\item for the $z>0.4$ sample, we defined a low mass range for $10.6<log({\cal M/M}_{\odot})<11$ (566 galaxies) 
and a high mass range for $log({\cal M/M}_{\odot})>11$ (365 galaxies).
\end{itemize}

\begin{figure}[t!]
\begin{center}
\includegraphics[width=0.48\textwidth]{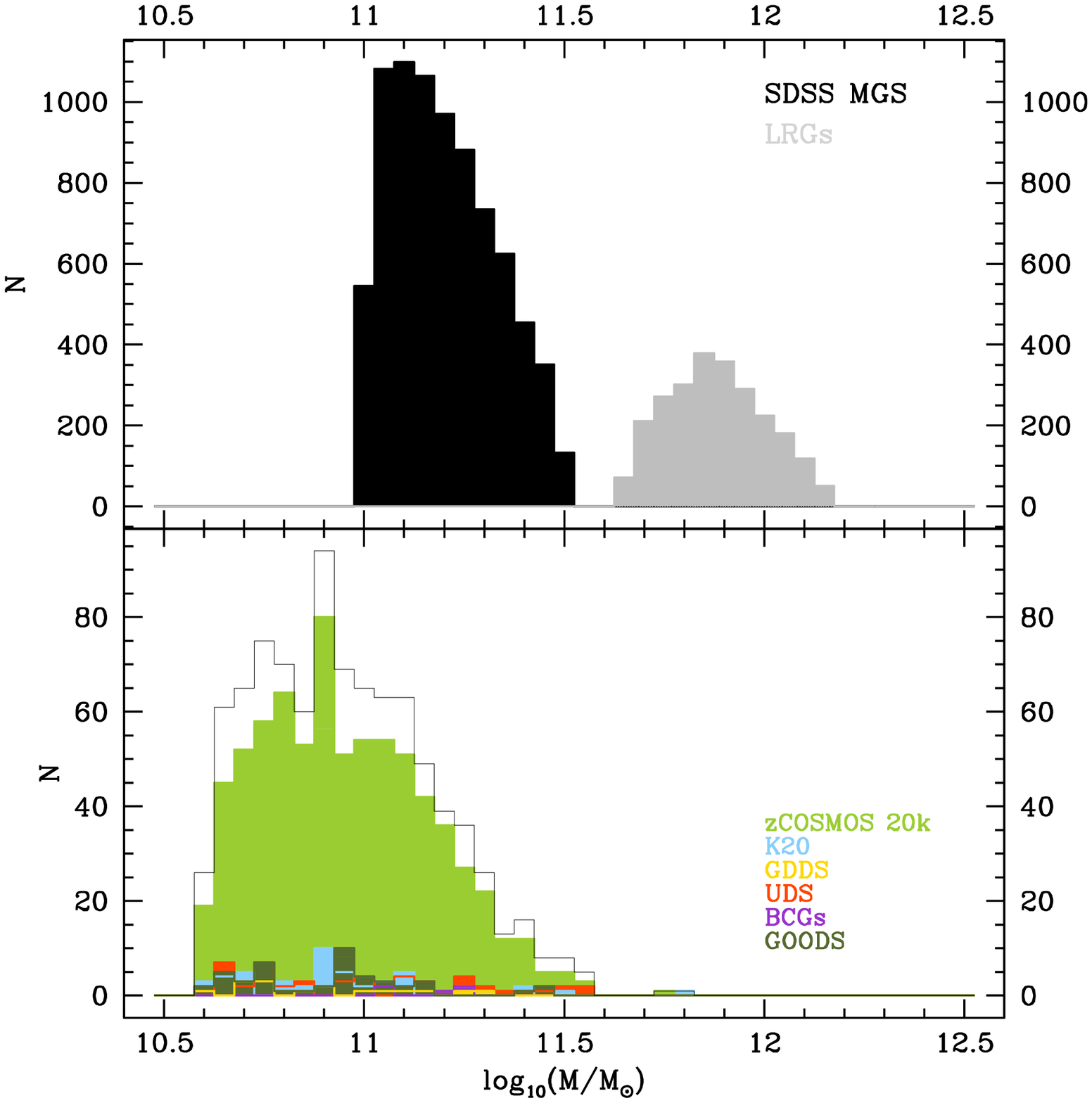}
\includegraphics[width=0.48\textwidth]{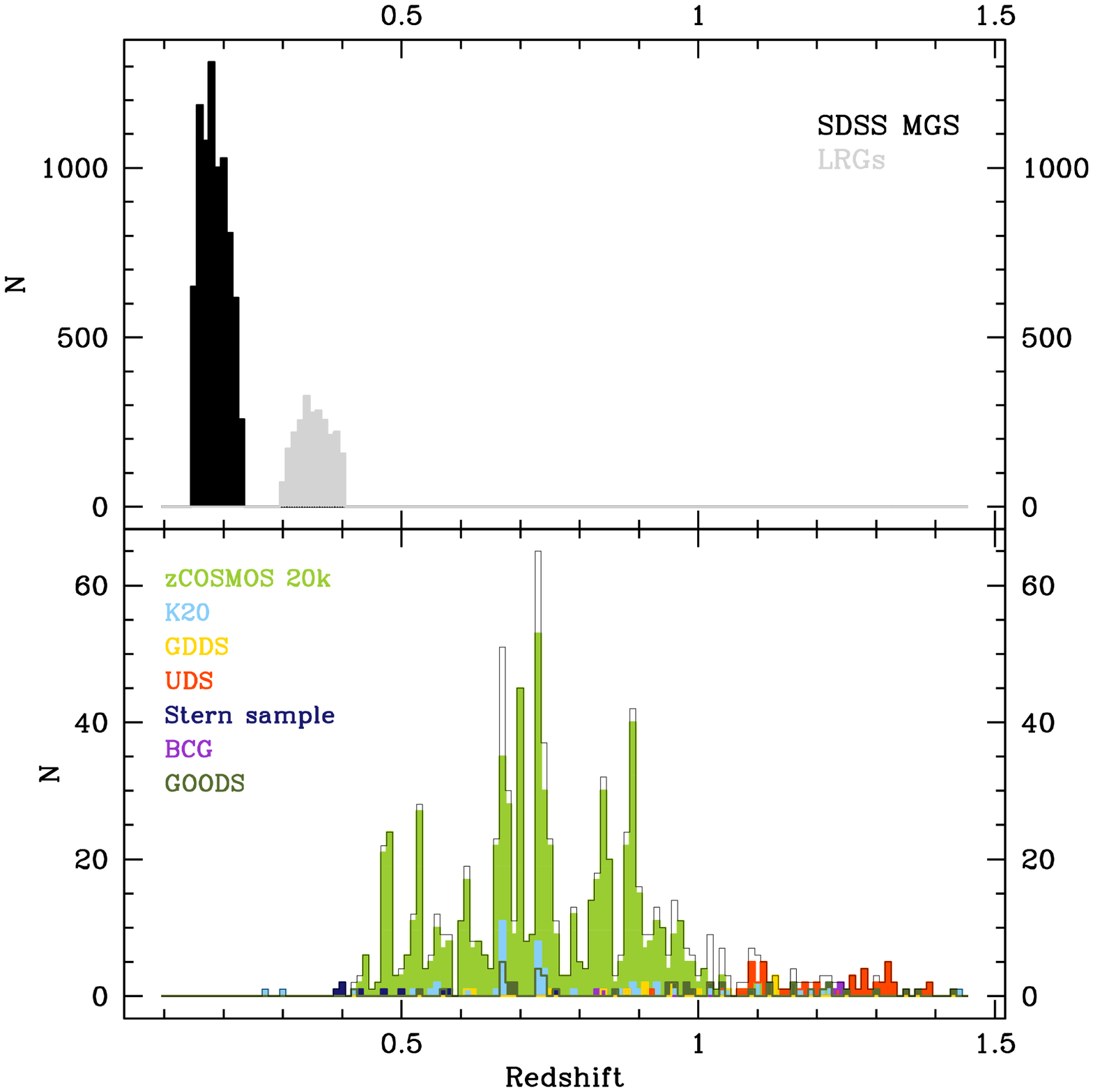}
\end{center}
\caption{Stellar mass histograms (left panel) and redshift distributions (right panel) of the ETG samples.
\label{fig:masshist}}
\end{figure}

\begin{figure}[b!]
\begin{center}
\includegraphics[width=0.85\textwidth]{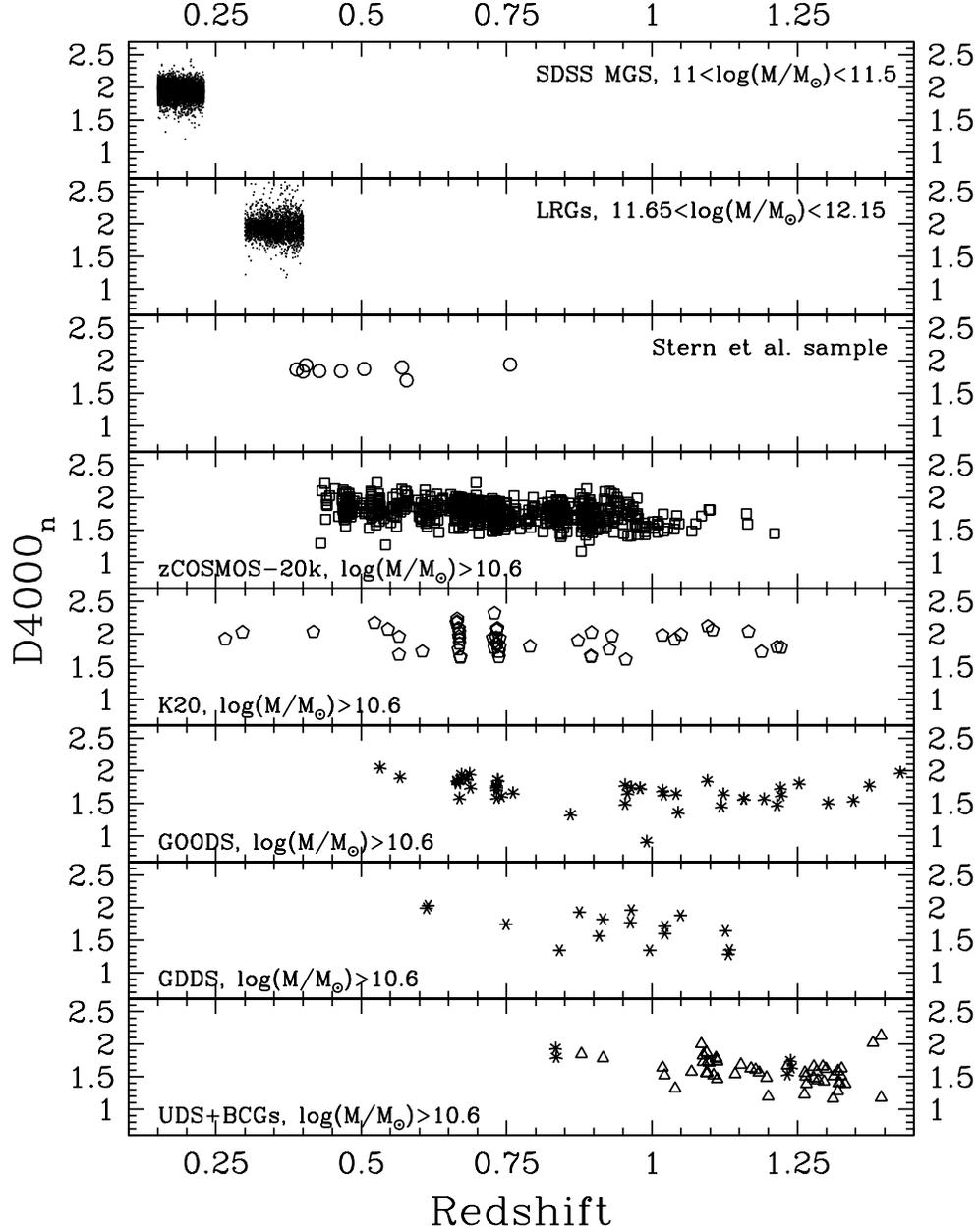}
\end{center}
\caption{$D4000_{n}-z$ plots for all the ETG samples separately.
\label{fig:D4000z-sep}}
\end{figure}

In this way, we ensure in each subsample a nearly constant mass as a function of redshift
and we avoid mixing the redshift evolution characteristic of different mass regimes.
In the sample at $z>0.4$, we decided to consider also the mass range $10.6<log({\cal M/M}_{\odot})<11$
to extend the $H(z)$ analysis with another mass bin, since the low statistics of this sample
does not allow us to further divide the mass bin with $log({\cal M/M}_{\odot})>11$. 
Figure \ref{fig:masshist} shows the redshift and mass 
distributions of the overall sample; the redshift distribution in particular 
shows clearly the presence of structures in the redshift range $0.45<z<1$. 
The $D4000_{n}-z$ plots for all the samples separately are shown in Fig. 
\ref{fig:D4000z-sep}. 

%%%%%%%%%%%%%%%%%%%%%%%%%%%%%%%%%%%%%%%%%%%%%%%%%%%%%%%%%%%%%%%%%%%%%%%%%%%%%%%%%%%%%%%%%%%
\section{The method: from spectra to ages}
\label{sec:D4000z}

The aim of this section is to illustrate the method with which we
used the 4000~\AA~break as a proxy of the stellar
age to estimate the expansion rate of the Universe, $H(z)$.

The 4000~\AA~break is a feature in galaxy spectra that was firstly introduced by Ref.~\cite{Bruzual1983}
as the ratio between the continuum flux densities in a red band (4050-4250~\AA) and a blue band (3750-3950~\AA) around
4000~\AA~restframe: 
\begin{equation}
D4000=\frac{(\lambda_{2}^{blue}-\lambda_{1}^{blue})\int_{\lambda_{1}^{red}}^{\lambda_{2}^{red}}F_{\nu}d\lambda}{(\lambda_{2}^{red}-\lambda_{1}^{red})\int_{\lambda_{1}^{blue}}^{\lambda_{2}^{blue}}F_{\nu}d\lambda}
\end{equation}
We decided to adopt the slightly different definition introduced by Ref.~\cite{Balogh1999} (hereafter $D4000_{n}$), 
where narrower bands (3850-3950~\AA~and 4000-4100~\AA) have been used
in order to be less sensitive to dust reddening.

The amplitude of this feature, due to metal absorption lines, depends on the age and metallicity of the stellar
population, as well as on the star formation history. However, within 
specific ranges of $D4000_{n}$, it has been demonstrated that 
its amplitude correlates linearly with the age of the 
galaxy and is weakly dependent on the star formation history if
the stellar population is old and passively evolving \cite{Moresco}.

The expansion rate is
\begin{equation}
H(z) = \frac{\dot a}{a} = \frac{-1}{1+z} \frac{dz}{dt},
\end{equation}
where $a(z)=1/(1+z)$.\\
M11 introduced an approximate linear 
relation between $D4000_{n}$ and galaxy age (at fixed metallicity):
\begin{equation}
D4000_{n}(Z, SFH)=A(Z,SFH)\cdot age+B(Z,SFH),
\label{eq:linD4000age}
\end{equation}
where $A(Z,SFH)$ (in units of Gyr$^{-1}$) is the conversion factor between age and 
$D4000_{n}$. This approximation has the considerable 
advantage that the relative $D4000_{n}$ evolution directly traces the age 
evolution of a population of galaxies:
\begin{equation}
\Delta D4000_{n}=A(Z,SFH)\cdot\Delta age.
\label{eq:lindD4000dage}
\end{equation}
The Hubble parameter $H(z)$ can, then, be rewritten as a function of the 
differential evolution of $D4000_{n}$:
\begin{equation}
H(z)=-\frac{A(Z,SFH)}{1+z}\frac{dz}{dD4000_{n}}.
\label{eq:HzD4000}
\end{equation}
Thus, to estimate $H(z)$ using the approach just discussed, it is 
therefore necessary to:
\begin{enumerate}
\item derive an observed $D4000_{n}-z$ mean relation, that will provide 
the quantity $dz/dD4000_{n}$;
\item calibrate the $D4000_{n}$--age relation with stellar population
synthesis models and therefore quantify the $A(Z,SFH)$ parameter;
\item estimate $H(z)$, verifying the robustness of the results against 
the adopted choice of binning, stellar population synthesis model, 
SFH, and metallicity.
\end{enumerate}

\subsection{The importance of the differential approach}
\label{sec:diff}
One of the main challenges for the study of galaxies as a function of redshift,
and even more in the analysis we are performing, is to ensure and validate the assumption 
that we are looking at the same population as a function of redshift, so that
their properties may be meaningfully compared at different redshifts.
Several issues may falsify this assumption. One of the most significant is ``progenitor-bias'', 
which refers to the issue that samples of ETGs at high redshift might be biased towards the oldest 
progenitors of present-day early-type galaxies \cite{Franx1996,vanDokkum2000}, therefore not
sampling the same population studied at intermediate redshifts. Another effect to be taken into account is the mass evolution
of ETGs as a function of redshift, which, when comparing galaxies of the same mass, may alter the shape
of the age$-z$ relation. Finally, one also has to take into account the fact that, lacking
a unique spectroscopic survey suitable to study ETGs evolution over a wide redshift range,
we are forced to merge together information
obtained from surveys with different selection criteria, photometry, mass 
estimates, systematics and so on.

However, it is fundamental to emphasize that our method relies
only on a differential measurement, since the estimate of $H(z)$ is based only on the measurement
of the quantity $dD4000/dz$. Therefore, each $H(z)$ point is obtained {\it not} by comparing
ETGs at $z\sim0$ with ETGs at $z\sim1$, so that the effects described above may play a
significant role, but instead comparing points close in redshift, with 
$\Delta z\sim0.04$ at $z<0.4$ and $\Delta z\sim0.3$ at $z>0.4$, as is discussed in the next section. 
Converted into terms of cosmic time, the previously quoted $\Delta z$ correspond to differences 
in cosmic time of $\sim500$ Myr for $z<0.4$ and of 
$\sim1.5$ Gyr at $z>0.4$, which are short times for potential effects due to merging or
mass evolution: this helps in mitigating the problem of the mass evolution and also of the progenitor-bias, 
which typically can affect studies where the properties of distant ETGs are 
directly compared to those of nearby ones. In Appendix \ref{sec:progbias} it is discussed how we checked
the reliability of our analysis against this effect, and a quantitative estimate
based on observational constraints is given, but it is as well discussed how, given the present errors,
these estimates have to be considered as upper limits, since our data are still compatible with not being biased by such an effect.

We also emphasize that for massive and passive ETGs, the measured
evolution in terms of mass and number density is observed to be less significant as compared to less massive galaxies.
Ref.~\cite{pozzetti} shows that ETGs with $log({\cal M/M}_{\odot})>11$ are compatible with no evolution in number density from $z\sim1$ 
to $z\sim0$, while for less massive ones ($10.7<log({\cal M/M}_{\odot})<11$) one observes an evolution roughly of a factor three over the same redshift
range. This redshift range corresponds to a range of cosmic times five times longer than the range of cosmic time covered by our differential estimates.
Ref.~\cite{Brammer2011} shows that the average mass in individual quiescent galaxies grows by a factor of
$\sim2$ from $z=2$ to $z=0$, spanning again a range of cosmic time $\sim7$ times longer than the 
range used in our analysis.

Moreover, we are also treating the different ETG samples (SDSS MGS, LRGs and ``$z>0.4$'' ETG samples)
separately, estimating $H(z)$ only within each sample. In this way, the differences 
between the various samples represent less than an issue, 
as long as the homogeneity of selection criteria, mass measurement and systematics treatment
is guaranteed within each sample and the redshift
evolution is estimated independently within each subsample. What is important is
 simply to ensure a uniform sampling where the differences are taken,
but a global uniform sample (e.g. in terms of mass ranges and absolute ages) is not required, as 
these differences would just produce systematic constant offsets 
that will drop out when evaluating a differential quantity.

\subsection{The observed $D4000_{n}-z$ relation}
\label{sec:D4000z_med}

The observed $D4000_{n}-z$ relation for the entire ETG sample 
is shown in the upper panel of Fig. \ref{fig:Hzcomparison}. Each orange 
and green point at $z<1.5$ indicates the median value of $D4000_{n}$ in a 
given redshift bin for a given mass range, green for the lower mass
range and orange for the higher mass range as described in section \ref{sec:sample}. The points
in the gray shaded area represent the single $D4000_{n}$ measurements of Ref.~\cite{Onodera2010,Ferreras2011,Kriek2009}.
The choice of the redshift bin width has been made taking into
account the difference between the SDSS MGS subsample ($z<0.3$), LRG
subsample ($0.3<z<0.4$), and the other subsamples ($z>0.4$).

Due to its high statistics, the adopted bin width for the SDSS MGS subsample
is $\Delta z=0.02$, with each bin including $\gtrsim1000$ galaxies for the low mass bin
and $\sim500$ for the high mass bin. For the LRGs subsample,
which have much lower statistics, we have used a wider binning 
to keep the number of galaxies per bin sufficiently high, with $\Delta z=0.05$
and each bin having $\gtrsim500$ galaxies both for the high and the low mass bins. 
For the other subsamples at $z>0.4$, an adaptive redshift binning 
centered around the structures present in the redshift distribution 
has been used (see Fig. \ref{fig:masshist}), with wider bins where there were no structures. 
In this way, it was also possible to obtain an almost constant number 
of galaxies per bin, with $N_{gal}\sim70$ for the low mass bin and 
$N_{gal}\sim50$ for the high mass bin. The size of the redshift bins is fundamental, 
since too wide a redshift binning will produce a poorly sampled 
$H(z)$, and too narrow a redshift binning will produce oscillations in 
the $D4000_{n}-z$ relation. 

In each redshift bin, the median $D4000_{n}$ was then derived,
separately for each mass regime. 
The associated errors are standard deviations on the median, 
defined as the {\it ``median absolute deviation''} (MAD, 
$\mathrm{MAD}=1.482\cdot \mathrm{median}(|D4000_{n}-\mathrm{median}(D4000_{n})|)$) 
divided by $\sqrt{N}$, i.e.
$\sigma_{med}(D4000_n)=\mathrm{MAD}/\sqrt{N}$ (see Ref.~ \cite{Hoaglin1983}). 

As discussed above, the $D4000_{n}$ is a feature that depends both on age and metallicity,
and less significantly on other parameters such as the assumed SFH, IMF or $\alpha-$enhancement 
($<10\%$, see Sect. \ref{sec:model} and Appendix \ref{sec:othersyst}). It is therefore important to obtain informations
about the metallicity of our ETGs.
For the SDSS MGS ETG sample, where the signal-to-noise ratio and the wavelength 
coverage of the spectra allows us to estimate the metallicity, 
the median metallicity $Z/Z_{\odot}$ and their errors 
$\sigma_{med}(Z/Z_{\odot})$ have been derived. On average, we find a slightly super-solar
metallicity, with a mean value close to $Z/Z_{\odot}=1.1$ almost constant, only slightly decreasing with redshift. 
For the higher-redshift samples, where metallicity estimates are not 
available, we made a conservative choice of assuming
a metallicity $Z/Z_{\odot}=1.1\pm0.1$, because this range largely contains 
all the possible median values of the metallicity found in our SDSS MGS sample,
also considering larger mass cuts.

The assumption to assign to ETGs at $z>0.3$ the same range
of metallicity observed in ETGs at $z<0.3$ is well justified
by the evolutionary scenario of the most massive ETGs; ETGs
should not show a significant metallicity evolution because
the vast majority of their stars were already formed at higher
redshifts ($z>2-3$) and most of their gas was consumed, 
hence not allowing significant changes of the metallicity with 
respect to $z\sim 0$ (e.g. see \cite{Carson2010}). This picture is also supported by direct 
measurements of solar to slightly super-solar metallicities in massive 
ETGs up to $z\sim1$ (e.g. see \cite{Ziegler2005,Martinez2011})
and also at higher redshifts ($z\sim2$, \cite{Toft2012}).
As a consistency check, in Fig. \ref{fig:stackspec} we overplotted to a
high-z ETGs stacked spectrum a BC03 spectrum with delayed $\tau$ SFH 
($\tau=0.1$ Gyr), solar metallicity and age of 2.5 Gyr; the model spectrum 
has been convolved at a velocity dispersion of 250 ${\rm km s^{-1}}$, typical 
of the ETGs considered. From the figure it is possible to notice the good agreement
between the model and the observed spectra.

Table \ref{tab:D4000med} shows, for each redshift bin, the median 
$D4000_{n}$ values and their errors, the median metallicity 
$Z/Z_{\odot}$ values and their errors, the median masses
and the number of galaxies analyzed. 

\begin{table}[t!]
\begin{center}
\begin{tabular}{lllllll}
\hline \hline
& & & low mass & & \\
\hline
$z$ & $D4000_{n}$ & $\sigma_{med}(D4000_n)$ & $Z/Z_{\odot}$* & $\sigma_{med}(Z/Z_{\odot})$& $log({\cal M/M}_{\odot})$ & \# gal\\
\hline
0.16 & 1.961 & 0.002 & 1.119 & 0.007 & $11.1$ & 1870\\
0.1796 & 1.944 & 0.002 & 1.075 & 0.008 & $11.12$ & 1543\\
0.1982 & 1.932 & 0.003 & 1.07 & 0.009 & $11.13$ & 1157\\
0.2164 & 1.921 & 0.004 & 1.066 & 0.01 & $11.14$ & 640\\
0.3291 & 1.936 & 0.004 & 1.1 & 0.1 & $11.79$ & 792\\
0.3705 & 1.912 & 0.006 & 1.1 & 0.1 & $11.81$ & 618\\
0.4712 & 1.85 & 0.02 & 1.1 & 0.1 & $10.83$ & 47\\
0.5313 & 1.84 & 0.01 & 1.1 & 0.1 & $10.86$ & 72\\
0.6624 & 1.82 & 0.01 & 1.1 & 0.1 & $10.81$ & 103\\
0.729 & 1.75 & 0.01 & 1.1 & 0.1 & $10.8$ & 118\\
0.8235 & 1.71 & 0.02 & 1.1 & 0.1 & $10.81$ & 87\\
0.8931 & 1.68 & 0.02 & 1.1 & 0.1 & $10.83$ & 73\\
1.02 & 1.64 & 0.03 & 1.1 & 0.1 & $10.83$ & 38\\
\hline
& & & high mass & & \\
\hline
$z$ & $D4000_{n}$ & $\sigma_{med}(D4000_n)$ & $Z/Z_{\odot}$* & $\sigma_{med}(Z/Z_{\odot})$& $log({\cal M/M}_{\odot})$ & \# gal\\
\hline
0.1589 & 1.992 & 0.003 & 1.21 & 0.01 & $11.33$ & 565\\
0.181 & 1.976 & 0.003 & 1.11 & 0.01 & $11.34$ & 682\\
0.1995 & 1.962 & 0.003 & 1.1 & 0.01 & $11.34$ & 864\\
0.2202 & 1.947 & 0.003 & 1.07 & 0.01 & $11.36$ & 622\\
0.3316 & 1.948 & 0.005 & 1.1 & 0.1 & $11.97$ & 388\\
0.3765 & 1.939 & 0.006 & 1.1 & 0.1 & $12$ & 661\\
0.4409 & 1.928 & 0.04 & 1.1 & 0.1 & $11.15$ & 13\\
0.5292 & 1.875 & 0.03 & 1.1 & 0.1 & $11.14$ & 34\\
0.6682 & 1.854 & 0.02 & 1.1 & 0.1 & $11.11$ & 62\\
0.7305 & 1.808 & 0.01 & 1.1 & 0.1 & $11.19$ & 67\\
0.8347 & 1.771 & 0.01 & 1.1 & 0.1 & $11.15$ & 46\\
0.9011 & 1.797 & 0.02 & 1.1 & 0.1 & $11.14$ & 72\\
1.022 & 1.71 & 0.03 & 1.1 & 0.1 & $11.16$ & 47\\
1.239 & 1.575 & 0.03 & 1.1 & 0.1 & $11.26$ & 24\\
\hline \hline
\end{tabular}
\caption{The median $D4000_{n}$, metallicity $Z/Z_{\odot}$ and $log({\cal M/M}_{\odot})$ as a 
function of redshift, and relative uncertainties ($\sigma_{med}(D4000_n, Z/Z_{\odot}) = \mathrm{MAD}/\sqrt{N}$, see text). $^{*}$ The metallicity has been estimated only for the 
SDSS MGS ETG sample, where the signal-to-noise ratio of the spectra was high enough; in the $z>0.3$ sample, a metallicity $Z/Z_{\odot}=1.1\pm0.1$ has been assumed.}
\label{tab:D4000med}
\end{center}
\end{table}

\subsection{The calibration of the $D4000_{n}$-age relation}
\label{sec:model}

This section describes how the relations between
$D4000_{n}$ and age were derived.
First, in order to mitigate uncertainties due the choice
of the stellar population synthesis model, we adopted two independent
libraries of synthetic spectra: the BC03 
models \cite{bc03} and the new MaStro models \cite{mastro}. 
These two models differ substantially for the stellar evolutionary models used to construct the isochrones, for the 
treatment of the termally pulsing asymptotic giant branch (TP-AGB) phase, and for the procedure used for computing 
the integrated spectra (e.g. see \cite{Maraston2005,Maraston2006} for more detailed reviews).
The two models are also based on independent libraries of stellar spectra, 
with MaStro using the latest MILES models \cite{Miles2011} and BC03 using STELIB \cite{Stelib2003}. 
The metallicities provided by the two models are $Z/Z_{\odot}=[0.4,1,2.5]$ 
for BC03 and $Z/Z_{\odot}=[0.5,1,2]$ for MaStro. The resolution of the 
two models is similar, 3~\AA~ across the wavelength range from 3200~\AA~ 
to 9500~\AA~ for BC03, and 2.54~\AA~ across the wavelength range 3525~\AA~ 
to 7500~\AA~ for MaStro. 

\begin{figure}[t!]
\begin{center}
\includegraphics[width=0.48\textwidth]{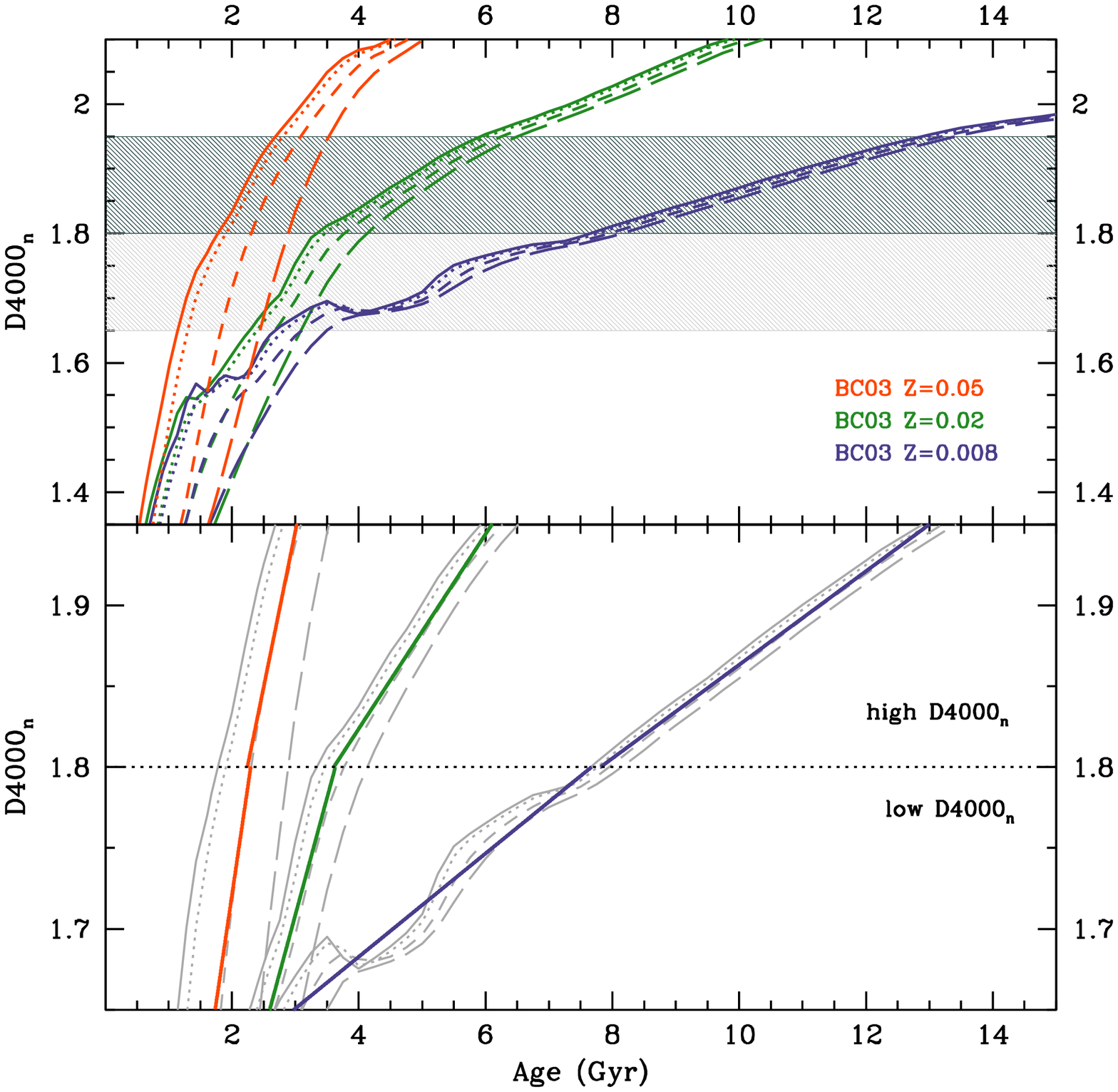}
\includegraphics[width=0.48\textwidth]{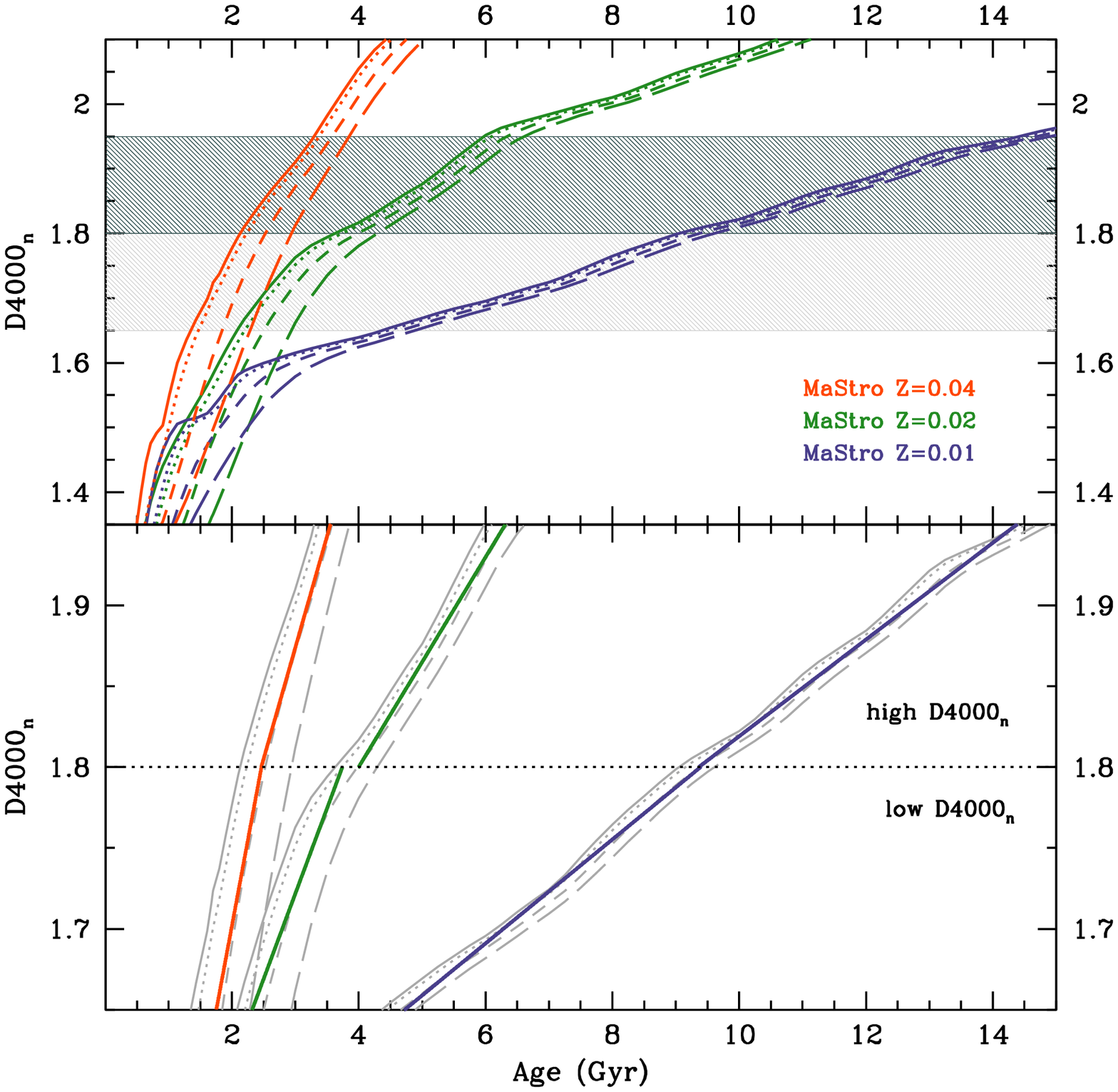}
\end{center}
\caption{$D4000_{n}$-age relation for BC03 models (left panel) and for MaStro models (right panel). In the upper panel the colored lines represent models with different metallicities, with super-solar metallicity in orange ($Z/Z_{\odot}=2.5$), solar in green, and sub-solar in blue ($Z/Z_{\odot}=0.4$); for each metallicity, the models are plotted with a continuous line for $\tau=0.05$ Gyr, a dotted line for $\tau=0.1$ Gyr, a dashed line for $\tau=0.2$ Gyr and a long-dashed line for $\tau=0.3$ Gyr. Different grey shaded areas represent the ranges of the {\it high $D4000_{n}$} and the {\it low $D4000_{n}$} regime. The lower panel shows a zoom of the area of interest, where the models are shown in gray and the colored lines are fits to the models. The dotted line divides the {\it high $D4000_{n}$} and the {\it low $D4000_{n}$} regimes.
\label{fig:D4000-ageBC03}}
\end{figure}

For both models, the $D4000_{n}$-age relations have been derived for four 
different SFHs, using an exponentially delayed $SFR(t)\propto 
t/\tau^{2}\,exp(-t/\tau)$ with $\tau=[0.05,0.1,0.2,0.3]$ Gyr.
The choice of the grid of $\tau$ has been done considering that the SFHs adopted 
must be compatible with the observed SEDs and spectra of passive ETGs. 
From analysis of the SDSS MGS ETG sample, Ref.~\cite{Moresco} 
found that the $\tau$ distribution presents a median value below 
0.2 Gyr for all of the mass subsamples, and that $\tau\leq 0.3$ Gyr is required
for the majority of the observed SEDs. The same applies to
galaxies at $z>0.3$; e.g. the SED-fit analysis of zCOSMOS 20k ETGs finds, allowing $\tau$ 
free up to 1 Gyr, a median value of $\tau=0.3$ Gyr. This result is consistent with
several other studies of ETGs \cite{Gobat2008,Longhetti2009,Rettura2010}, for which 
it has been found that the majority of massive field and cluster ETGs 
formed the bulk of their stellar mass at $z\gtrsim2$ over 
short (i.e. $\tau<0.1-0.3$ Gyrs) star formation time-scales. 

Since we want to ensure the validity of the linear approximation
in the relations between $D4000_{n}$ and the stellar age, we found 
that it is convenient to divide them into two regimes of $D4000_{n}$ 
values which span the $D4000_{n}$ range observed in the ETG spectra 
of our sample:
\begin{itemize}
\item the {\it low $D4000_{n}$} is defined for $1.65<D4000_{n}<1.8$ 
\item the {\it high $D4000_{n}$} is defined for $1.8<D4000_{n}<1.95$ 
\end{itemize}
The $D4000_{n}$-age relation from BC03 and MaStro models derived
independently for these two regimes are shown in Fig. \ref{fig:D4000-ageBC03}. 
These figures also show
the dependence on metallicity and SFH (for the adopted range of $\tau$). 

\begin{figure}
\begin{center}
\includegraphics[width=0.48\textwidth]{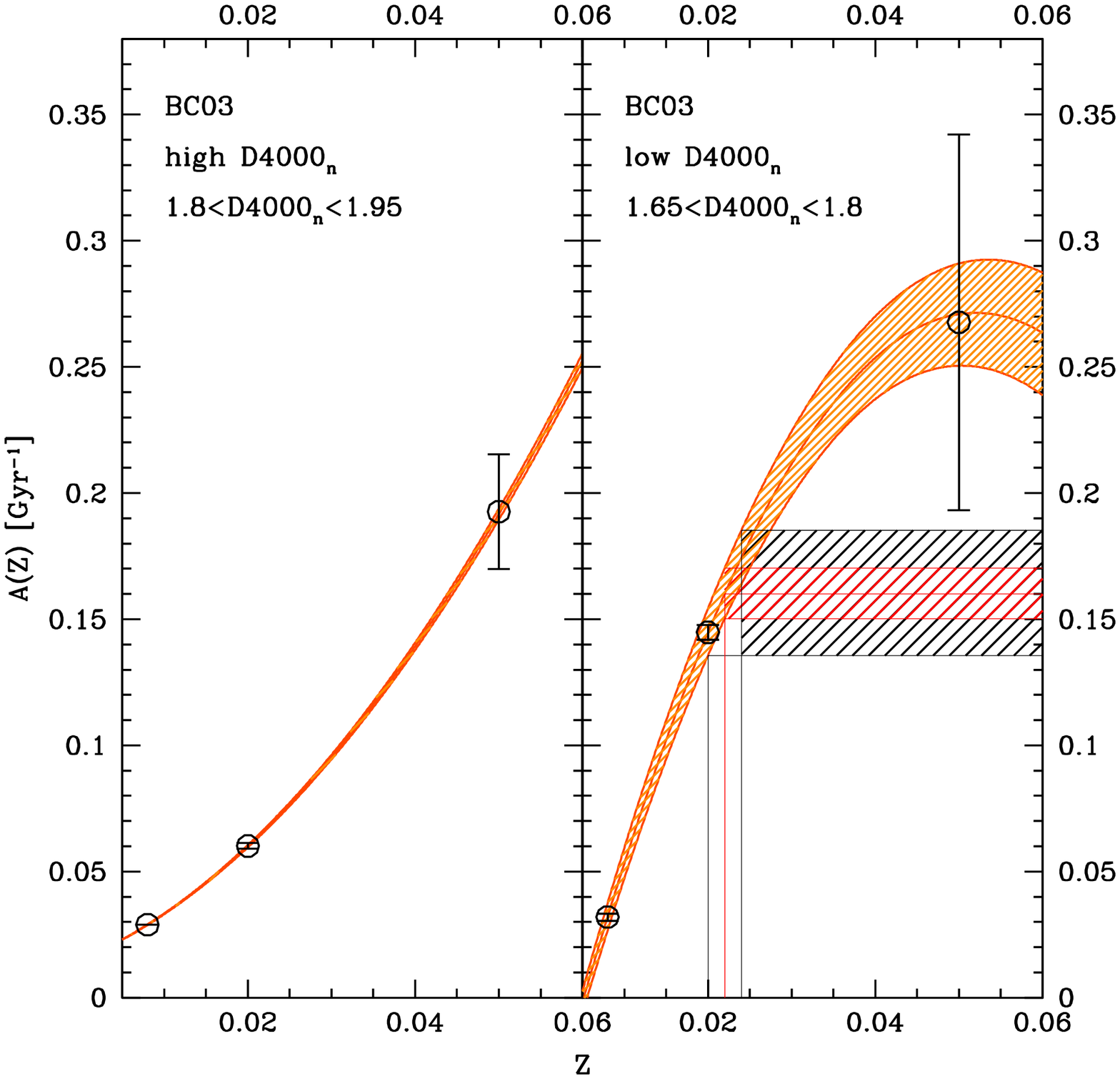}
\includegraphics[width=0.48\textwidth]{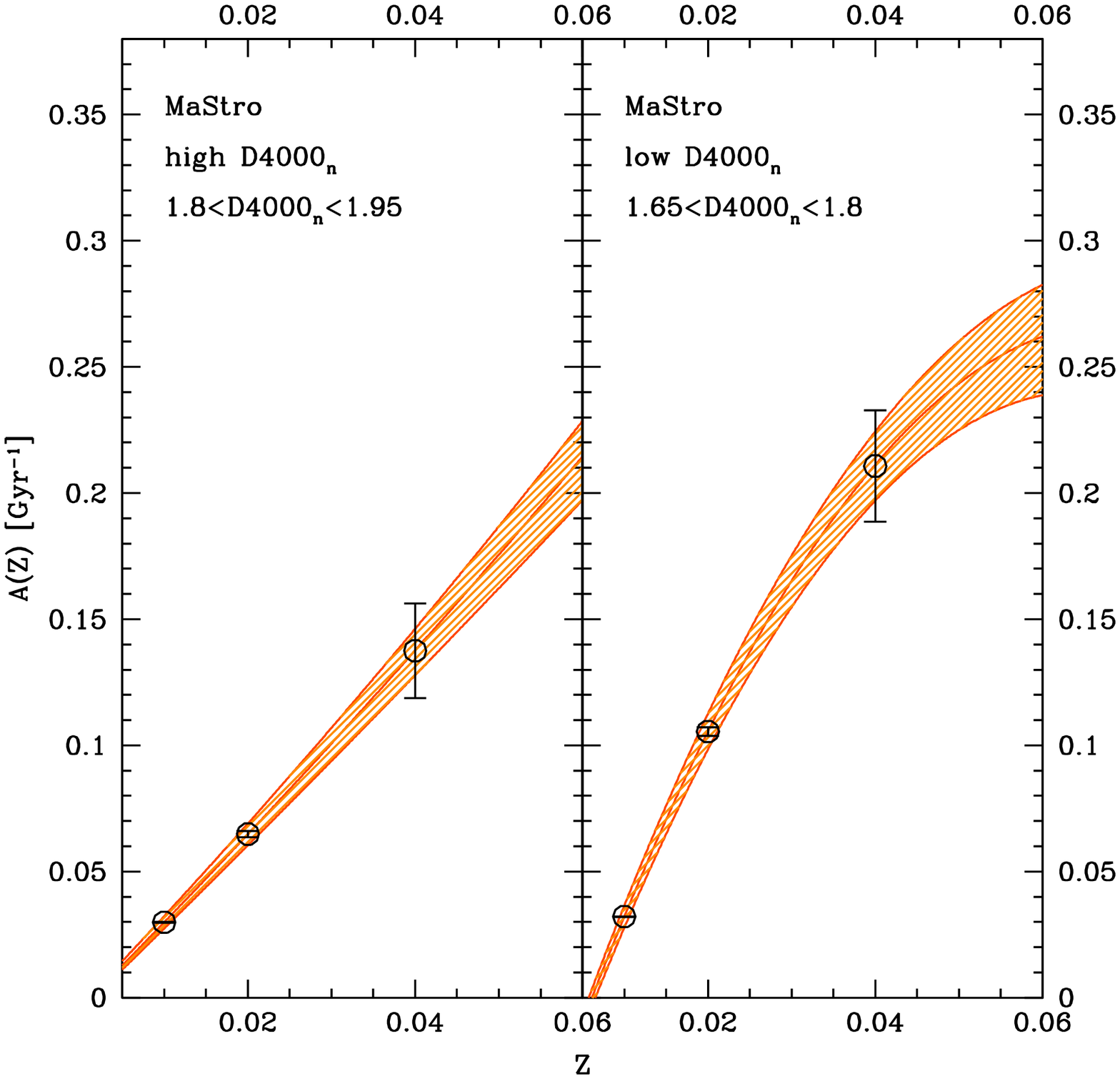}
\end{center}
\caption{$A=f(Z)$ relation (in units of [Gyr$^{-1}$], see eq. \ref{eq:linD4000age}) for BC03 models (left panel) and for MaStro models (right panel). The black points represent the mean slope obtained for the different metallicities in each $D4000_{n}$ regime, and the orange shaded area shows the interpolation and the associated $1\sigma$ error. In the left panel, the red shaded area represents the contribution to the range of allowed $A(Z)$ due to SFHs assumption, while the black shaded area show the total range of $A(Z)$ when is considered both SFH and metallicity uncertainty (see Sect.~\ref{sec:Hzerror}). 
\label{fig:ZslopeBC03}}
\end{figure}

A best-fit slope $A(Z,SFH)$ has been estimated for each $D4000_{n}$ regime.
The linear approximation is valid and accurate for all the regimes and 
metallicities, having in the case of BC03 models linear correlation 
coefficients $>0.996$ with a mean value of $0.9987\pm0.0004$ 
for the {\it ``high $D4000_{n}$''} regime and $>0.966$ with a mean 
value of $0.990\pm0.003$ for the {\it ``low $D4000_{n}$''} regime. In the 
case of MaStro models the linear correlation coefficients are always $>0.997$ 
with a mean value of $0.9984\pm0.0003$ for the {\it ``high $D4000_{n}$''} regime 
and $>0.986$ with a mean value of $0.995\pm0.001$ for 
the {\it ``high $D4000_{n}$''} regime.

In the two $D4000_n$ ranges, a mean slope for each metallicity $A(Z)$ 
has then been obtained by averaging the slopes $A(Z,SFH)$ obtained for the 
four SFHs, and considering the dispersion between these measurements
as the associated error. Those values are listed in Tab. \ref{tab:Zslope}, and 
shown in Fig. \ref{fig:ZslopeBC03}.

For a fixed metallicity, the similarity of the individual and mean
slopes, as well as the small errors associated with the mean slopes 
$A(Z)$, confirm that, for $\tau\leq 0.3$ Gyr, the dependence of 
the $D4000_{n}$-age relation on the SFH is negligible with respect to 
the dependence on metallicity. We also checked the effects of extending the grid 
up to $\tau=0.6$ Gyr, and we found that the results on the slopes
$A(Z)$ are still consistent within $1\sigma$. If we adopt a different 
functional shape of the SFH (e.g. a declining exponential 
$SFR(t)\propto 1/\tau\,exp(-t/\tau)$) the results on $A(Z)$ are 
compatible within $0.5\sigma$ with the ones obtained with the 
exponentially delayed SFHs. 

\begin{table}[b!]
\begin{center}
\begin{tabular}{llll}
\hline \hline
& & high $D4000_{n}$ & low $D4000_{n}$\\
\hline
BC03 & $A(Z/Z_{\odot}=0.4)$&$0.02893\pm0.00004$&$0.032\pm0.001$\\
BC03 & $A(Z/Z_{\odot}=1)$&$0.060\pm 0.001$&$0.145\pm0.003$\\
BC03 & $A(Z/Z_{\odot}=2.5)$&$0.193\pm0.002$&$0.27\pm0.07$\\
\hline \hline
& & high $D4000_{n}$ & low $D4000_{n}$\\
\hline
MaStro & $A(Z/Z_{\odot}=0.5)$&$0.0299\pm0.0002$&$0.0321\pm0.0001$\\
MaStro & $A(Z/Z_{\odot}=1)$&$0.065\pm 0.001$&$0.106\pm0.002$\\
MaStro & $A(Z/Z_{\odot}=2)$&$0.138\pm0.02$&$0.21\pm0.02$\\
\hline \hline
\end{tabular}
\caption{Mean slopes $A(Z)$ of the $D4000_{n}$-age relation (in units of [Gyr$^{-1}$], see Eq. 
\ref{eq:linD4000age}) for the BC03 and MaStro models with different 
metallicities and for the two $D4000_{n}$ regimes. The quoted errors are 
the dispersion between the slopes evaluated for different choices of SFH.}
\label{tab:Zslope}
\end{center}
\end{table}

Following the approach of M11, for each $D4000_{n}$ regime the obtained 
values of the slopes (quoted in Tab. \ref{tab:Zslope}) have been 
interpolated with a quadratic function, shown in Fig. 
\ref{fig:ZslopeBC03}. When the metallicity 
is known, these relations allow one to associate the correct $A(Z)$ parameter 
to each given metallicity.

%%%%%%%%%%%%%%%%%%%%%%%%%%%%%%%%%%%%%%%%%%%%%%%%%%%%%%%%%%%%%%%%%%%%%%%%%%%%%%%%%%%%%%%%%%%
\section{The estimate of $H(z)$}
\label{sec:Hzevaluation}

The estimate of $H(z)$ is based 
on two quantities (eq. \ref{eq:HzD4000}): 
\begin{itemize}
\item the relative $D4000_{n}$ evolution as a function of redshift 
\item the parameter $A(Z)$ for a given metallicity
\end{itemize}
The quantity $dz/dD4000_{n}$ has been calculated directly from the median 
$D4000_{n}-z$ relation shown in the upper panel of Fig.~\ref{fig:Hzcomparison}, 
estimating the $\Delta D4000_{n}$ between the $i$-th and the $(i+N)$-th point
for each mass bin. The choice of $N$ was the result of a trade-off analysis between 
two competing effects: on the one hand, we want $N$ as small as 
possible to maximize the number of $H(z)$ measurements, on the other hand,
$N$ has to be large enough in order to have a $D4000_{n}$ evolution 
larger than the statistical scatter present in the data, in order to ensure an 
unbiased estimate of $H(z)$. For the SDSS MGS sample, where we have four median $D4000_{n}$ points,
we chose $N=2$, to have two estimates of $H(z)$ mutually independent;
for the LRGs sample, having only two $D4000_{n}$ points, $N$ has to be equal to one.
For the $z>0.4$ sample, the analysis indicated $N=3$ as the 
best compromise, being the smallest value ensuring a redshift evolution
larger than the scatter in the data.
This choice provides a redshift leverage between ${i}$-th and
$(i+N)$-th point of $\Delta z\approx0.04$ for the SDSS MGS sample
and the LRGs sample, and of $\Delta z\approx0.3$ for $z\geq0.4$ samples,
which corresponds to a difference in cosmic time of $\sim500$ Myr and of 
$\sim1.5$ Gyr respectively.

For the high-z sample, the differential $D4000_{n}$ evolution has been estimated between
a median value of $D4000_{n}$ and the $D4000_{n}$ value of the individual galaxies, i.e. 
respectively the third last point with the galaxy
of Ref.~\cite{Onodera2010}, the second last point with the galaxy of Ref.~\cite{Ferreras2011},
and the last point with the galaxy of Ref.~\cite{Kriek2009}.

For the choice of $A(Z)$ needed in eq. \ref{eq:HzD4000}, as described
in the previous section, the $D4000_{n}-z$ relation has been divided into 
two parts: if the median $D4000_{n}$ value is greater than 1.8, then 
we extrapolate the $A(Z)$ values using the {\it high $D4000_{n}$} regime, 
while if the median $D4000_{n}$ value is smaller than 1.8, then $A(Z)$ 
values obtained from the {\it low $D4000_{n}$} regime have been used. 
In the case in which the two points used to estimate $\Delta D4000_{n}$ 
lie one in the first regime and one in the other, a median relation between 
the {\it high $D4000_{n}$} and the {\it low $D4000_{n}$} regimes 
has been adopted.

For the metallicity, we distinguish two cases: the SDSS MGS subsample
at $z<0.3$, where $Z$ is known for each ETG, and the rest of the
sample at $z>0.3$ where the metallicity is unknown due to the
limited signal-to-noise ratio and/or the limited wavelength coverage
of the spectra. For the SDSS MGS ETGs, we adopted the observed median 
metallicity for each redshift bin, as quoted in Tab. \ref{tab:D4000med}. 
For the ETGs at $z>0.3$, as discussed in section \ref{sec:D4000z_med}, 
we assumed a median metallicity $Z/Z_{\odot}=1.1\pm0.1$. 
Thus, this metallicity range $\Delta Z$ enters as an additional 
uncertainty in the $H(z)$ error budget (see next section). 

Given these metallicities, $A(Z)$ is obtained from the interpolated 
$A=f(Z)$ relation described in section \ref{sec:model} and shown in 
Fig. \ref{fig:ZslopeBC03}. 
The detailed procedure is described in the following section, where we 
also discuss how the metallicity uncertainties (as well as the 
SFH uncertainties) are treated in our error estimate.

The measurements of $H(z)$ have proven to be extremely robust even changing 
between completely different stellar population synthesis models: performing the
analysis separately with the MaStro and the BC03 model, the values obtained are 
in agreement with a mean difference of $0.5\pm0.4\sigma$, except for the last point where 
there is a difference of $1.6\sigma$. The results will be discussed in section \ref{sec:results},
and the lower panel of Fig. \ref{fig:Hzcomparison} shows the comparison of the $H(z)$ measurements 
for the two models.

\subsection{$H(z)$ error budget}
\label{sec:Hzerror}

There are two main sources of error in the $H(z)$ estimate: 
a statistical error related to the computation 
of $dz/dD4000_{n}$, which depends on the median $D4000_{n}$ 
errors, and a systematic error related to the estimate of $A(Z)$, which 
depends on the metallicity range spanned by the data, on the SFH
assumptions, as well on the adopted stellar population synthesis model. 

\paragraph{Statistical error}
In the previous section we described the method used to estimate the
relative $D4000_{n}$ evolution, and the statistical error $\sigma_{stat}$ 
has been obtained with standard error propagation.
\paragraph{Systematic error}
To estimate the systematic uncertainty, we consider two effects.\\
\begin{enumerate}
\item {\bf SFH contribution} The spread of the $A=f(Z)$ relation due to SFH assumption causes, at the typical 
metallicity of our ETGs ($Z/Z_{\odot}\sim1.1$), an error on the estimate of $A(Z)$
of about 13\% in the high and low $D4000_{n}$ regimes for the MaStro models,
and of $\sim$2\% and $\sim$13\% respectively in the high $D4000_{n}$ and low $D4000_{n}$ 
regimes for the BC03 models. This uncertainty is
shown, as an example, as the red shaded region in Fig. \ref{fig:ZslopeBC03}.\\
\item {\bf Metallicity contribution} To consider also the uncertainty
due to metallicity, we have associated to each $H(z)$
measurement three possible values of $A(Z)$, representing the minimum, 
the median, and the maximum value of $A(Z)$ allowed by the metallicity range. As 
shown in Fig. \ref{fig:ZslopeBC03}, the minimum value of $A(Z)$ has been
obtained by considering the lowest metallicity in a given redshift bin (i.e. 
$Z/Z_{\odot}-\sigma_{med}(Z/Z_{\odot})$, see Tab. \ref{tab:D4000med}) 
and using the lowest possible $A=f(Z)$ interpolated relation; 
the median $A(Z)$ value has been obtained by considering 
the median metallicity (i.e. $Z/Z_{\odot}$) and the best-fit interpolated relation; 
the maximum $A(Z)$ value has been obtained by considering the highest metallicity
in a given redshift bin (i.e. $Z/Z_{\odot}+\sigma_{med}(Z/Z_{\odot})$) and 
using the highest possible $A=f(Z)$ interpolated relation. In this way, to the previous
uncertainty due to the SFHs assumption, we add an uncertainty on $A(Z)$ that
depends specifically on the range of metallicity considered (e.g. for the range of metallicity of the $z>0.3$
samples, it is $\sim$20\% and $\sim$22\% for the high and low $D4000_{n}$ 
regimes with MaStro, and $\sim$20\% and $\sim$19\% for the high 
and low $D4000_{n}$ regimes with BC03). The total uncertainty
due to the SFHs and metallicity is represented by the black shaded region of 
Fig. \ref{fig:ZslopeBC03}.
\end{enumerate}
For each of the three values of $A(Z)$ described in point (2.) we estimate a value of $H(z)$, and the dispersion between these measurements quantifies 
the systematic error $\sigma_{syst}$.\\

Each $H(z)$ measurement is, then, estimated as the weighted mean of the $H(z)$ values obtained
with the three values of $A(Z)$, and the total error on $H(z)$ has been obtained by 
summing in quadrature the statistical error $\sigma_{stat}$ and the systematic error 
$\sigma_{syst}$.

%%%%%%%%%%%%%%%%%%%%%%%%%%%%%%%%%%%%%%%%%%%%%%%%%%%%%%%%%%%%%%%%%%%%%%%%%%%%%%%%%%%%%%%%%%%

\section{Results and cosmological implications}
\label{sec:results}

The Hubble parameter $H(z)$ has been estimated as described in section \ref{sec:Hzevaluation} 
separately for the SDSS MG sample, the LRGs sample, and the samples at $z>0.4$;
for both BC03 and MaStro models, the estimates have been obtained 
for each of the mass ranges described in section \ref{sec:sample}. We find that
the $H(z)$ estimates obtained for the two different mass ranges show good agreement, therefore
providing strong evidence that this approach is not dependent on the chosen mass range.
Since these estimates are statistically independent,
they have been averaged using a weighted mean of the $H(z)$ points at the same redshift,
using as weights the corresponding error of each measurement.

\begin{figure}[t!]
\includegraphics[width=0.9\textwidth]{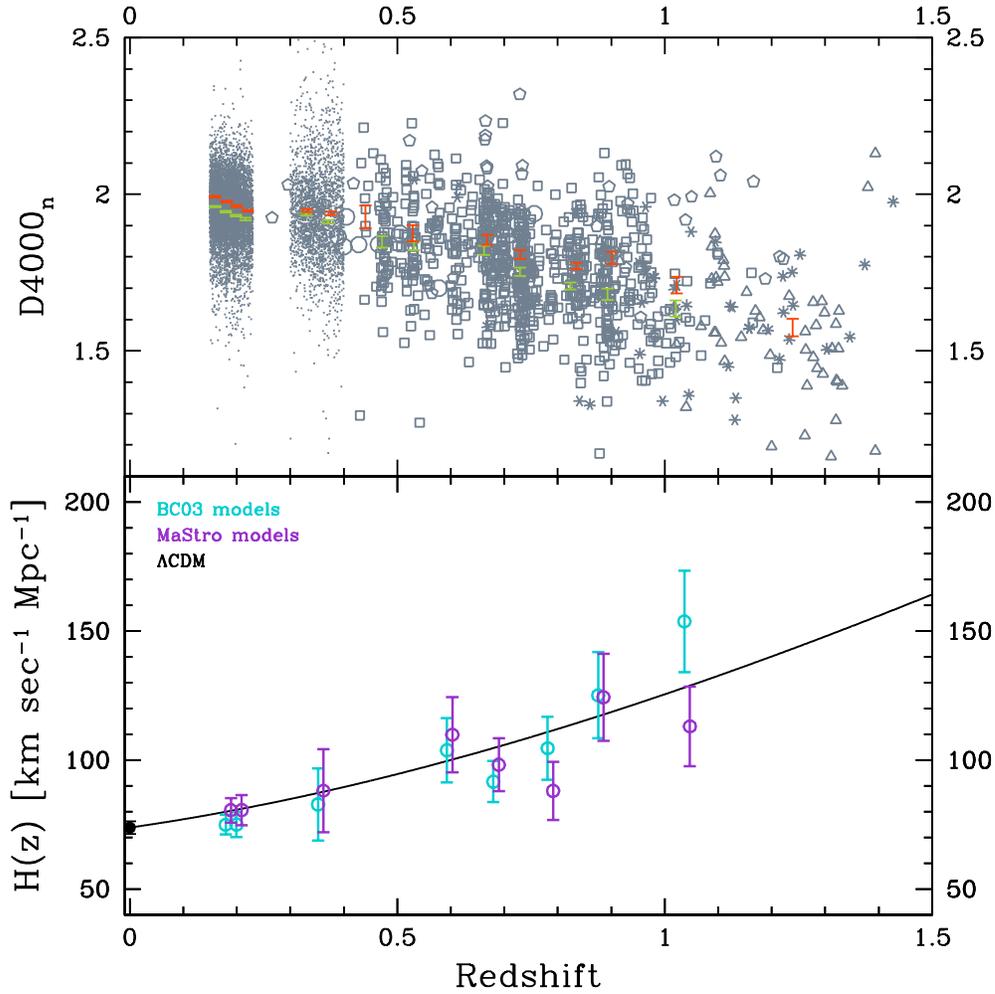}
\caption{The $D4000_n$-redshift relation and $H(z)$ 
measurements. The $D4000_n$ feature measured from different surveys is 
shown in the upper panel as a function of redshift. Grey, green and orange 
symbols indicate the D4000$_n$ values for individual galaxies and its median 
value in each redshift bin, respectively for the low and the high mass subsamples (as described in Sect.~\ref{sec:sample}). 
The $H(z)$ values have been estimated between the $i$-th and the $(i+N)$-th
points as described in Sect. \ref{sec:Hzevaluation}, not to be biased by the statistical scatter of the data;
in the lower panel are shown the results relative to BC03 (in cyan) and to MaStro (in violet) stellar population synthesis 
model and their $1\sigma$ uncertainties; $H(z)$ estimates obtained with MaStro models have been slightly offset in redshift for
the sake of clarity. The solid point at $z=0$ represents the measurement of Ref.~\cite{Riess}.
As a comparison, we also show the $H(z)$ relation for the $\Lambda$CDM model, assuming a flat WMAP 7-years Universe [56], with $\Omega_{m}=0.27$, $\Omega_{\Lambda}=0.73$ and $H_{0}=73.8\mathrm{\;km\;s^{-1}Mpc^{-1}}$.
\label{fig:Hzcomparison}}
\end{figure}

\begin{figure}[t!]
\includegraphics[width=0.9\textwidth]{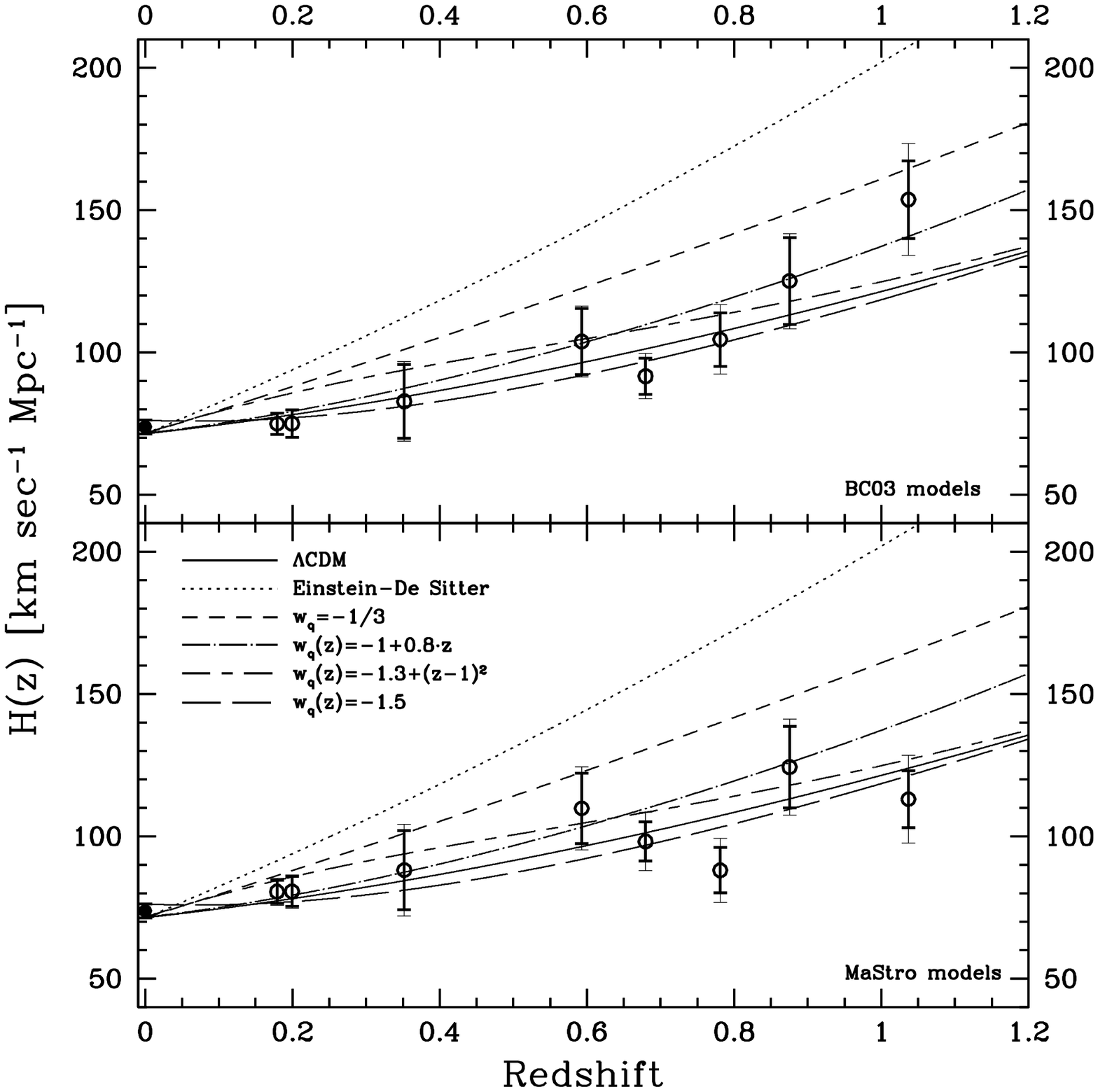}
\caption{$H(z)$ measurements and comparison with theoretical models. 
The $H(z)$ measurements are shown in the upper and in the lower panel assuming respectively 
a BC03 and a MaStro stellar population synthesis model. 
The thicker errorbars represent the $1\sigma$ statistical errors, while the thinner show the total 
$1\sigma$ errors. The solid point at $z=0$ represents the measurement of Ref.~\cite{Riess}.
The curves are different theoretical $H(z)$ relations, assuming a flat WMAP 7-years Universe, with $\Omega_{m}=0.27$, 
$\Omega_{\Lambda}=0.73$, and $H_{0}$ as the best fit value in the range $H_{0}=73.8\pm2.4\mathrm{\;km\;s^{-1}Mpc^{-1}}$.
\label{fig:HzBC03}}
\end{figure}

The results, obtained using BC03 and MaStro stellar population synthesis models, are compared in the bottom panel of Fig. \ref{fig:Hzcomparison},
separately shown in the upper and lower panels of Fig. \ref{fig:HzBC03}, and reported in Tab. \ref{tab:HzBC03}. 
In these tables we report for each measurement the statistical error $\sigma_{stat}$ 
and the systematic error $\sigma_{syst}$; we also reported the total error, estimated by summing in quadrature 
the statistical and the systematic error, with its absolute and percentage value.
The local value of $H_{0}$ shown refers to the recent measure of Ref.~\cite{Riess}, with $H_{0}=73.8\pm2.4\mathrm{\;km\;s^{-1}Mpc^{-1}}$
obtained including both statistical and systematic errors. After averaging the $H(z)$ estimates obtained from the two 
mass ranges, we obtained a fractional error on $H(z)$ at $z<0.3$ of the order of $\sim5\%$, which is 
close to the value estimated locally in M11; 
we recall that Ref.~\cite{Stern} achieved an accuracy of only
$\sim15\%$ at $z<0.3$. Furthermore, our derived $H(z)$ values are fully 
compatible with $\Lambda$CDM, constraining the expansion rate 
very tightly. 

At higher redshifts, the error increases because 
of the smaller number of observed galaxies in the samples, but still provide typical accuracy
of (12-13)\% on $H(z)$ up to $z\sim1$, which improves by a factor $2-3$ current measurements in this redshift range. The 
redshift range $0.5<z<1$ is critical to disentangle many different cosmologies,
as can be seen by the upper and lower panels of Fig. \ref{fig:HzBC03}.

The comparison between the measurements obtained with BC03 and MaStro models is shown in the lower panel of Fig. 
\ref{fig:Hzcomparison}. Even though the models are based on completely different stellar libraries and evolutionary synthesis codes,
the $H(z)$ measurements show very good agreement: we conclude that our results are basically independent 
of the assumed stellar population synthesis model.

In particular, the present work goes beyond the present state of the art
\cite{Stern} in the following aspects: the use of different samples which give consistent results, ensuring that the
sample selection does not introduce any systematic error; the analysis performed with two different stellar models, which provide 
compatible results; the use of the $D4000$ feature, which is less model dependent and more robust than the approach of e.g. 
Ref.~\cite{Stern}; the control of the mass dependence systematics; the homogeneous coverage of the full redshift interval, and 
particularly the cosmic time between 5 and 8 Gyr ago, which is the crucial cosmic time when the Universe
changes from deceleration to acceleration; the higher accuracy across the entire redshift range (5--12\%, including systematic errors).

In the upper and lower panels of Fig. \ref{fig:HzBC03} we also show different theoretical $H(z)$ relations. Given that:
\begin{equation}
H(z)=H_{0}(1+z)^{3/2}\left[\Omega_{m,0}+\Omega_{\Lambda,0}exp\left(3\int_{0}^{z}\frac{w_{q}(z')}{1+z'}dz'\right)\right]^{1/2}
\end{equation}
we explored six different scenarios:
\begin{itemize}
\item $w_{q}(z)=-1$ (i.e. $\Lambda$CDM)
\item $\Omega_{\Lambda,0}=0$, $\Omega_{m,0}=1$ (i.e. Einstein - de Sitter, EdS hereafter)
\item $w_{q}(z)=-1/3$
\item $w_{q}(z)=-1.5$
\item $w_{q}(z)=-1+0.8\cdot z$
\item $w_{q}(z)=-1.3+(z-1)^{2}$
\end{itemize}

\begin{figure}[b!]
\begin{center}
\includegraphics[width=0.9\textwidth]{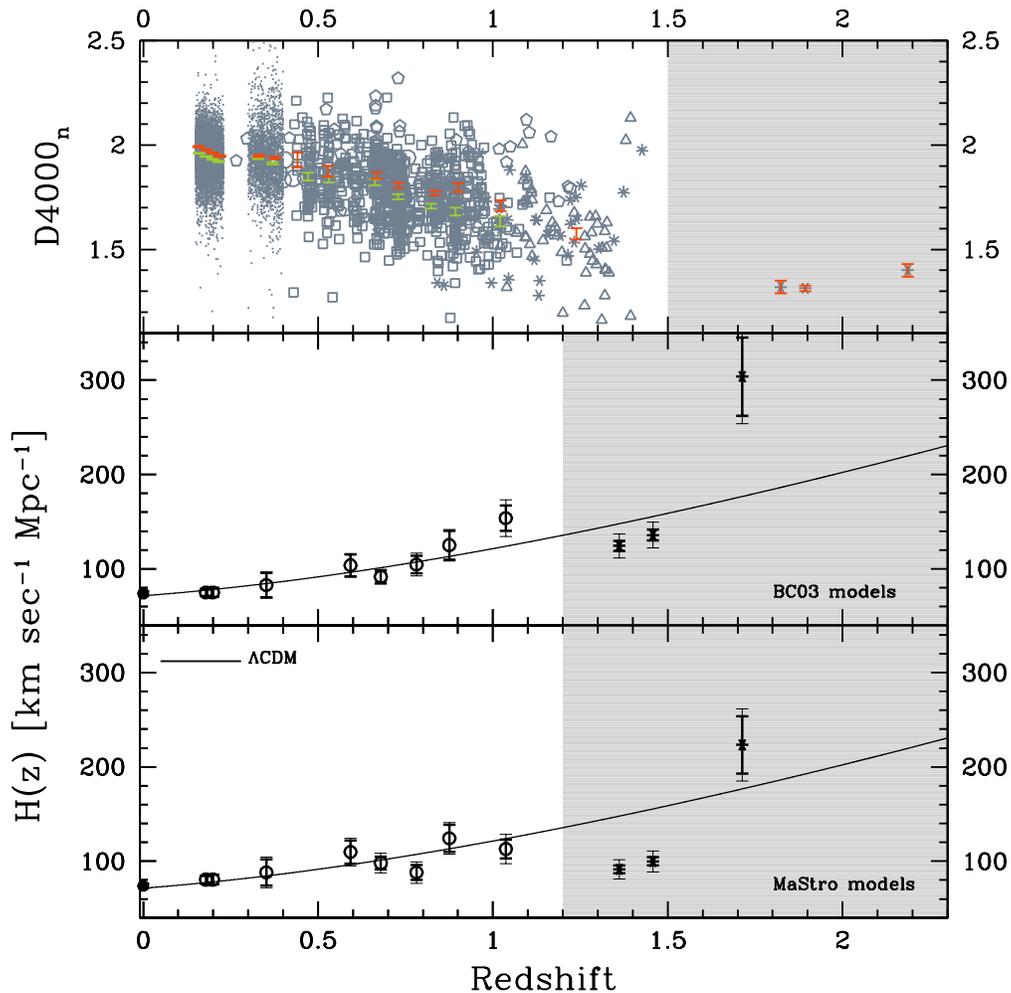}
\caption{The $D4000_n$-redshift relation and $H(z)$ 
measurements including the ``high-z'' sample. In the upper panel, 
the three starred points in the gray shaded area at $z>1.5$ correspond to the three individual galaxies at $1.8<z<2.2$ \cite{Onodera2010,Ferreras2011,Kriek2009}. 
In the middle and bottom panels are shown the $H(z)$ estimates, respectively for BC03 and MaStro stellar population
synthesis models. The three starred symbols in the gray area and the corresponding $H(z)$ estimates 
are not used in the $\chi^{2}$ comparison
with the models, because they are based only on three individual galaxies.
The curve represents the theoretical $H(z)$ relation for $\Lambda$CDM model, assuming a flat WMAP 7-years Universe [56], with $\Omega_{m}=0.27$, 
$\Omega_{\Lambda}=0.73$, and $H_{0}=73.8\pm2.4\mathrm{\;km\;s^{-1}Mpc^{-1}}$.
\label{fig:HzMastro}}
\end{center}
\end{figure}

The first two models have been chosen to see how our measurements compare to the standard EdS and $\Lambda$CDM model. 
The third model corresponds to the one discriminating between an accelerating and a non-accelerating Universe. We then considered 
three quintessence models (for a detailed discussion about theoretical models, see also Refs.~\cite{Peebles,Amendola}): the first one 
assumes a constant $w_{q}=-1.5$, and the other two are designed in such a way that they provide the same luminosity distance as 
$\Lambda$CDM model at the 1\% level. Thus it would be nearly impossible to distinguish them from $\Lambda$CDM with integrated 
measurements from standard candles, angular BAO or gravitational lensing measurements. The linear and quadratic model are 
somewhat ad-hoc, but they can be accommodated in more physically motivated models of quintessence. The models assume a flat Universe, 
$\Omega_{m,0}=0.27$, and $\Omega_{\Lambda,0}=0.73$ as pointed out by the latest WMAP 7-years results \cite{Komatsu2011}; the 
Hubble constant $H_{0}$ has been chosen as the value that minimizes the chi squared with respect to the data points, assuming 
$H_{0}$ in the range allowed by Ref.~\cite{Riess}, $H_{0}=73.8\pm2.4\mathrm{\;km\;s^{-1}Mpc^{-1}}$.

\begin{table}[t!]
\begin{center}
\begin{tabular}{||c||ccc|cc||ccc|cc||}
\hline
& \multicolumn{5}{c||}{BC03 models} & \multicolumn{5}{|c||}{MaStro models} \\
$z$ & $H(z)$ & $\sigma_{stat}$ & $\sigma_{syst}$ & $\sigma_{tot}$ & \% error & $H(z)$ & $\sigma_{stat}$ & $\sigma_{syst}$ & $\sigma_{tot}$ & \% error\\
\hline
0.1791 & 75 & 3.8 & 0.5 & 4 & $5\%$ & 81 & 4.1 & 2.5 & 5 & $6\%$\\
0.1993 & 75 & 4.9 & 0.6 & 5 & $7\%$ & 81 & 5.2 & 2.6 & 6 & $7\%$\\
0.3519 & 83 & 13 & 4.8 & 14 & $17\%$ & 88 & 13.9 & 7.9 & 16 & $18\%$\\\
0.5929 & 104 & 11.6 & 4.5 & 13 & $12\%$ & 110 & 12.3 & 7.5 & 15 & $13\%$\\
0.6797 & 92 & 6.4 & 4.3 & 8 & $9\%$ & 98 & 6.8 & 7.1 & 10 & $11\%$\\
0.7812 & 105 & 9.4 & 6.1 & 12 & $12\%$ & 88 & 8 & 7.4 & 11 & $13\%$\\
0.8754 & 125 & 15.3 & 6 & 17 & $13\%$ & 124 & 14.3 & 8.7 & 17 & $14\%$ \\
1.037 & 154 & 13.6 & 14.9 & 20 & $13\%$ & 113 & 10.1 & 11.7 & 15 & $14\%$\\
\hline
\end{tabular}
\caption{$H(z)$ measurements (in units of [$\mathrm{km\,s^{-1}Mpc^{-1}}$]) and their errors; the columns in the middle report
the relative contribution of statistical and systematic errors, and the last ones the total error (estimated by summing in quadrature $\sigma_{stat}$ and $\sigma_{syst}$). These values have been estimated respectively with BC03 and MaStro stellar population synthesis models. This dataset can be downloaded at the address 
http://www.astronomia.unibo.it/Astronomia/Ricerca/ Progetti+e+attivita/cosmic\_chronometers.htm (alternatively http://start.at/cosmicchronometers).}
\label{tab:HzBC03}
\end{center}
\end{table}

This represents a direct measurement of $H(z)$ without assuming any cosmological model. 
The observed Hubble parameter $H(z)$ has been compared with the theoretical relation
with a standard $\chi^{2}$ formalism. These results allow us to discard an EdS model at more than $7\sigma$, independent of 
the assumed stellar population synthesis model. From a purely observational point of view, 
it gives a direct $6\sigma$ evidence of an accelerated expansion with both BC03 and MaStro 
models, confirming the results obtained with other probes (e.g. \cite{riess_de,perlmutter}). Concerning the 
other models, the one that best fits the data is $\Lambda$CDM for both BC03 and MaStro, 
while BC03 measurements show some tension with the model with $w_{q}(z)=-1.3+(z-1)^{2}$
(at $\sim2\sigma$), and MaStro measurements show some tension with the model with 
$w_{q}(z)=-1+0.8\cdot z$ (at $\sim1.5\sigma$).

The comparison with evolving $w_{q}$ shows that in principle this method has the possibility to 
discriminate models that produce the same luminosity distance, which would not be possible using 
SNe, even with a future SN survey. We enphasize that in order to obtain these 
measurements, we used no single dedicated survey. This means that if a survey may provide 
at $z>0.3$ a statistic comparable to the one obtained in the SDSS MGS, it would in theory be possible 
to constrain $H(z)$ at the $\sim5\%$ level up to $z\sim1$, allowing the possibility to distinguish 
between dark-energy models with evolving $w_{q}(z)$ from $\Lambda$CDM.

The points plotted in the gray region of Fig. \ref{fig:HzMastro} represent the $H(z)$ estimates obtained with the ``high-z'' sample. 
As explained in Sect. \ref{sec:Hzevaluation}, lacking enough statistics at those redshifts, these $H(z)$ measurements have been
obtained by estimating the differential evolution between the last median points of the $D4000_{n}-z$ relation up to $z\sim1.4$
and single $D4000_{n}$ measurements: therefore they have not been used in the following for the comparison
with the theoretical model. Here we just want to show how it is possible to extend this method up to much higher redshifts,
and that future measurements of $D4000_{n}$ at high redshifts for ETGs will open the possibility 
to extend this approach up to $z\sim2$. For example, Euclid will identify spectroscopically the rarest and 
most massive quiescent galaxies (${\cal M/M}_{\odot}>4\cdot10^{11}$) at $z>1.8$ \cite{redbook}, providing a large spectroscopic sample
of high redshift ETGs, which will allow a much more detailed investigation of quintessence models.
Pushing the ``cosmic-chronometers'' approach to higher redshifts, will however require a more detailed 
study of the systematics. Moving to $z>1.5$ will mean to select galaxies which are closer to their redshift of
formation, and therefore it will become more important to establish their detailed SFH; also the treatment of the
progenitor-bias will be a more delicate matter. A good way to deal with this
issue, providing that enough statistics is available, would be to select the oldest galaxy at each redshift, 
to probe exactly the upper and redder envelope of the age-z relation.
Moreover, at higher redshifts (hence at younger ages) the theoretical stellar population synthesis models start to differ more, as testified by the 
larger differences found from the test with the ETGs at $z>1.8$ (see the $H(z)$ values at $z>1.2$ in Fig. \ref{fig:HzMastro}); therefore a better 
understanding of which can be the best model to be used will become important too. 
Given the issues to be faced just described, a priority to extend this method should be to ensure enough statistics, to reduce at minimum the
statistical errors and to establish the reliability and robustness of having selected really the oldest galaxies at each redshift, from which depend the H(z) measurements. 
In this respect, analyzing the Baryon Oscillation Spectroscopic Survey (BOSS, \cite{boss}) would be a good step forward at intermediate redshifts, providing spectra and redshifts of $\sim1.5$ 
million luminous galaxies to $z=0.7$.

\begin{figure}[t!]
\begin{center}
\includegraphics[angle=-90,width=0.9\textwidth]{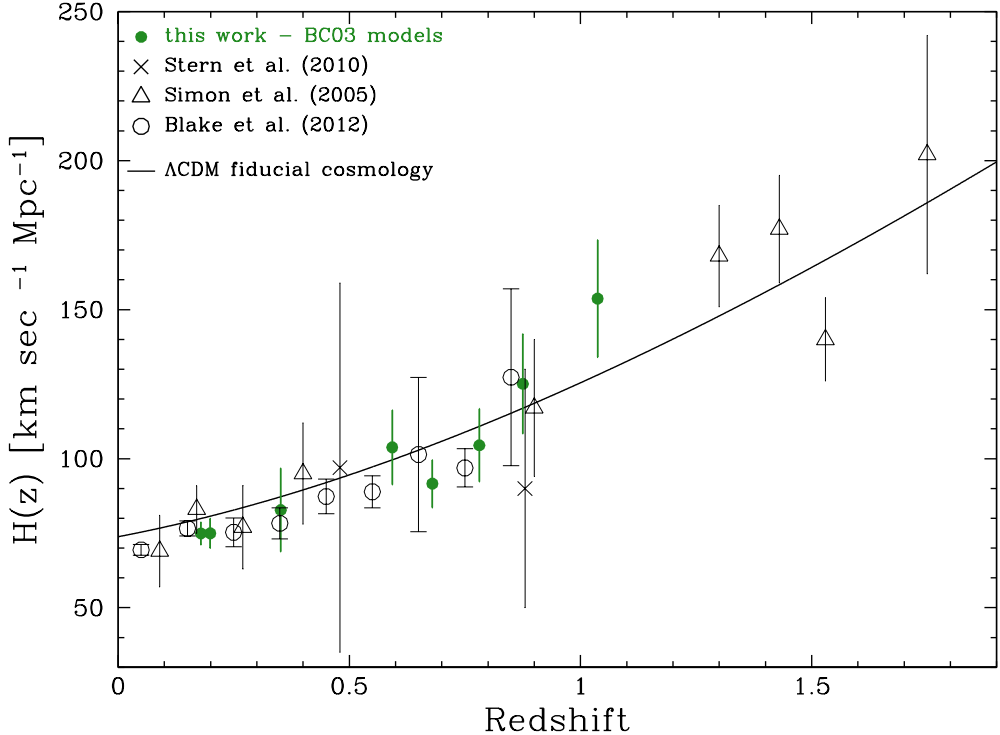}
\caption{Comparison of various $H(z)$ measurements. The green points are taken from this work 
(see Tab. \ref{tab:HzBC03}), the crosses from Ref. \cite{Stern}, the open triangles from Ref. \cite{Simon2005},
and the open dots from Ref. \cite{Blake2012}.
The $\Lambda$CDM model assumes a flat WMAP 7-years Universe, with $\Omega_{m}=0.27$, 
$\Omega_{\Lambda}=0.73$, and $H_{0}=73.8\pm2.4\mathrm{\;km\;s^{-1}Mpc^{-1}}$.
\label{fig:Hzall}}
\end{center}
\end{figure}

\section{Comparison with other $H(z)$ measurements}
\label{sec:comparison}

We compared our $H(z)$ measurements with data available in the literature \cite{Stern,Simon2005,Blake2012} obtained with other methods. 
The first two references use the ``cosmic chronometers'' approach to estimate $H(z)$ up to $z\sim1.8$, and are therefore useful to check the agreement
with the $H(z)$ evolution expected from our data; it is worthwhile to remember that these measurements estimate the differential ages of the oldest ETGs at each redshift in 
various surveys, and therefore do not suffer by definition of the progenitor-bias effect. Ref. \cite{Blake2012} uses instead a combination of measurements of the baryon 
acoustic peak and the Alcock-Paczynski distortion from galaxy clustering in the WiggleZ Dark Energy Survey, along with other baryon acoustic oscillation and 
distant supernovae datasets to determine the evolution of the Hubble parameter with a Monte Carlo Markov Chain technique up to $z\sim0.9$; in this case, 
having a precision better than 7\% in most redshift bins, these measurements are important to check the correctness of our data. The comparison between our dataset and the 
other Hubble parameter estimates is shown in Fig. \ref{fig:Hzall}, and we find a remarkable agreement between the different measurements: the $H(z)$ extrapolated at higher redshift from 
our data is in good agreement with the measurements of Refs. \cite{Simon2005,Stern}, and the comparison with the measurements of Ref. \cite{Blake2012} (in the case of overlapping 
redshift range) is always well within the $1\sigma$ errorbars.

\section{Conclusions}
We have analyzed the $D4000_{n}-z$ relation for a large sample of ETGs
extracted from different spectroscopic surveys; several selection criteria have
been combined in order to obtain a reliable sample comprising the most massive and
passive ETGs in the redshift range $0.15<z<1.42$. Due to the differences between the
various surveys, SDSS MGS ETGs, LRG ETGs and ``$z>0.4$'' ETGs have been studied separately,
and each subsample has been further divided into two mass subsamples, since in Ref. \cite{Moresco}
it has been shown a dependence of the $D4000_{n}-z$ relation on mass (i.e. downsizing).
In this way we obtained 7943 SDSS MGS ETGs with masses $10^{11}<{\cal M/M}_{\odot}<10^{11.5}$,
2459 LRG ETGs with masses $10^{11.65}<{\cal M/M}_{\odot}<10^{12.15}$, and 922 ``$z>0.4$'' ETGs
with masses $10^{10.6}<{\cal M/M}_{\odot}<10^{11.5}$.

For each subsample, a median $D4000_{n}-z$ relation has been estimated, and used to evaluate
the differential $D4000_{n}$ evolution as a function of redshift. In each subsample,
the couple of points used to evaluate the difference have been chosen as the
best compromise between a small redshift leverage, to provide the maximum number
of $H(z)$ measurements, and a long redshift leverage, not to be biased in the $\Delta D4000_{n}$
estimate by the statistical scatter present in the data.
In this way, we compare points separated by $\sim500$ Myr in cosmic time for SDSS MGS and LRG ETGs,
and $\sim1.5$ Gyr for ``$z>0.4$'' ETGs, considerably mitigating the possible effects of mass evolution
of our sample. Such effects are most onerous for analysis comparing $z\sim0$ to $z\sim1$ ETGs. (e.g. progenitor-bias).

We have studied the theoretical $D4000_{n}$-age relation for two different stellar population synthesis models,
BC03 \cite{bc03} and MaStro \cite{mastro}, to estimate the conversion factor $A(Z,SFH)$ between 
$\Delta D4000_{n}$ and $\Delta$age. This parameter, calibrated on the measured stellar metallicity for SDSS MGS ETGs
and on a extrapolated metallicity for LRG ETGs and ``$z>0.4$'' ETGs, have been estimated for different choices of SFHs,
and an averaged $A(Z)$ has been thus obtained. The effect of metallicity and SFHs, which may affect the
$H(z)$ estimate, have been studied and considered as systematic sources of error into the global
error budget of $H(z)$.

Finally, we estimated $H(z)$ separately for each mass subsample of SDSS MGS, LRGs and ``$z>0.4$'' ETGs. Since 
the results show a good agreement between the two mass subsamples, we averaged those estimate, to provide a 
single $H(z)$ estimate at each redshift which is mass independent. This analysis has been performed with both BC03 and 
MaStro models, and the results are in agreement with a mean difference of $0.5\pm0.4\sigma$, except for the last point where 
there is a difference of $1.6\sigma$, witnessing the robustness of the results against changes of stellar population synthesis
model.

We provide 8 new measurements of the observational Hubble parameter, mapping homogeneously the redshift
range $0.15<z<1.1$. At low redshift, we obtain for the first time a precision comparable to recent estimates of
the Hubble constant (5-6\% at $z\sim0.2$); at higher redshift, the precision decreases due to the decrease of statistics,
but still the errorbars are at $<13$\% up to $z\sim1.1$, considering both statistic and systematic errors.
We also show the possibility to extend our analysis up to much higher redshift by analyzing the $D4000_{n}$ of 
three high redshift ETGs at $z\sim1.7-2.2$; however, since in this case the $H(z)$ measurement is based only
on a single $D4000_{n}$ measurement, we do not use these points to constrain cosmological models.

Comparing these measurements with theoretical $H(z)$ relations, we obtain strong (6$\sigma$) evidence of
the accelerated expansion of the Universe, confirming the results obtained with other probes \cite{riess_de,perlmutter}.
An EdS Universe is ruled out at 7$\sigma$, and the model that best fits our data is $\Lambda$CDM.
The comparison with three quintessence models, built to be undistinguishable with respect to $\Lambda$CDM model at 
the 1\% level, show the capability of this approach to constrain different cosmologies.

Studying the $D4000_{n}$ of three high-z ETGs ($z>1.8$) we show the possibility to extend this approach constraining
$H(z)$ up to $z\sim2$. Given new upcoming spectroscopic surveys of ETGs (e.g. Euclid \cite{redbook} and BOSS \cite{boss}), 
the ``cosmic chronometer'' approach may represent a new complementary cosmological probe to place stringent constraints 
on the Dark Energy Equation of State parameter $w$ and its potential evolution with cosmic time. 

\section*{Acknowledgments}
MM and AC acknowledge contracts ASI-Uni Bologna-Astronomy 
Dept. Euclid-NIS I/039/10/0, and PRIN MIUR ``Dark energy and cosmology with large galaxy surveys''.
JSD acknowledges the support of the European Research Council through an Advanced Grant, and the 
support of the Royal Society via a Wolfson Research Merit award. The work of DS was carried out at the 
Jet Propulsion Laboratory, California Institute of Technology, under a contract with NASA. LV acknowledges the support of 
grant FP7 ERC- IDEAS Phys.LSS 240117.\\
This analysis is partly based on zCOSMOS observations, carried out using the Very Large Telescope 
at the ESO Paranal Observatory under Programme ID: LP175.A-0839, and partly based on SDSS observations.
Funding for the SDSS and SDSS-II has been provided by the Alfred P. Sloan Foundation, the Participating 
Institutions, the National Science Foundation, the U.S. Department of Energy, the National Aeronautics and 
Space Administration, the Japanese Monbukagakusho, the Max Planck Society, and the Higher Education 
Funding Council for England. The SDSS Web Site is http://www.sdss.org/. The SDSS is managed by the 
Astrophysical Research Consortium for the Participating Institutions. The Participating Institutions are the 
American Museum of Natural History, Astrophysical Institute Potsdam, University of Basel, University of Cambridge, 
Case Western Reserve University, University of Chicago, Drexel University, Fermilab, the Institute for Advanced Study, 
the Japan Participation Group, Johns Hopkins University, the Joint Institute for Nuclear Astrophysics, the Kavli 
Institute for Particle Astrophysics and Cosmology, the Korean Scientist Group, the Chinese Academy of Sciences 
(LAMOST), Los Alamos National Laboratory, the Max-Planck-Institute for Astronomy (MPIA), the Max-Planck-Institute 
for Astrophysics (MPA), New Mexico State University, Ohio State University, University of Pittsburgh, University of 
Portsmouth, Princeton University, the United States Naval Observatory, and the University of Washington.
The UDS spectroscopic data were obtained as part of the ESO Large Programme 180.A-0776 (PI: Almaini).\\
We would like to thank the anonymous referee for the useful comments and suggestions, which helped to improve the paper.
We are also grateful to M. Onodera and I. Ferreras for providing the $D4000_{n}$ measurements of their ETG \cite{Onodera2010,Ferreras2011},
J. Brinchmann for the help providing the matched SDSS-2MASS catalog, B. Hoyle for helping concerning SDSS spectra,
R. Tojeiro for the assistance about the VESPA catalog and C. Maraston for helpful discussion.

\appendix
\section{Other sources of systematic uncertainty}
\label{sec:othersyst}
There are other possible sources of uncertainty that may affect this analysis: the ``progenitor-bias'' effect, the choice of initial mass function (IMF)
adopted in the stellar population synthesis model studied, and the effect of the $\alpha-$enhancement, for which massive
ETGs have been found to higher ratios of alpha elements to iron than Milky Way-like galaxies. All these issues will be addressed separately
in the following sections.

\subsection{Estimating the effect of the progenitor-bias on $H(z)$}
\label{sec:progbias}

The ``progenitor-bias'' is an effect to be taken into account when two samples of ETGs at low- and high-redshifts are compared.
It has been firstly introduced by Refs. \cite{Franx1996,vanDokkum2000}, who pointed out that such comparison results biased
if the progenitors of the youngest ETGs at low redshift drop out of the sample at high redshift.

\begin{figure}[b!]
\begin{center}
\includegraphics[angle=-90,width=0.9\textwidth]{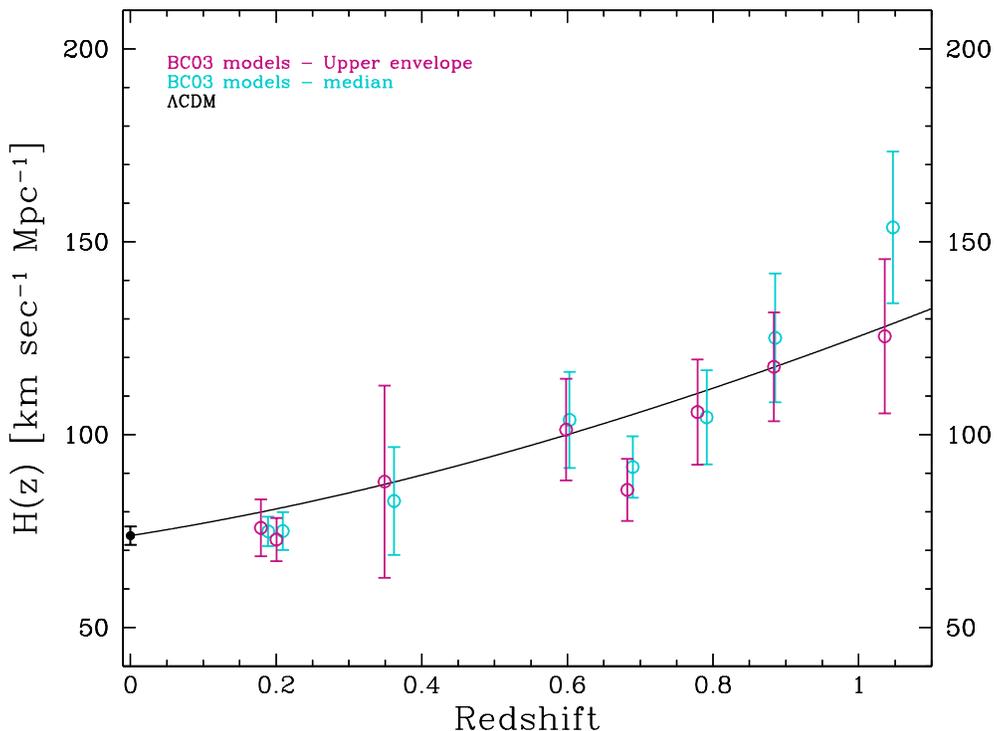}
\caption{Comparison of the $H(z)$ as measured from the median $D4000_{n}-z$ relations in different
mass subsamples and from the upper envelope. The $H(z)$ measurements shown assume 
BC03 stellar population synthesis model. The solid point at $z=0$ represents the measurement of Ref.~\cite{Riess}.
The $\Lambda$CDM model assumes a flat WMAP 7-years Universe, with $\Omega_{m}=0.27$, 
$\Omega_{\Lambda}=0.73$, and $H_{0}=73.8\pm2.4\mathrm{\;km\;s^{-1}Mpc^{-1}}$.
\label{fig:HzUE}}
\end{center}
\end{figure}

\begin{table}[t!]
\begin{center}
\begin{tabular}{|c|cc|cc|}
\hline
& \multicolumn{2}{|c|}{median} & \multicolumn{2}{|c|}{upper envelope}\\
$z$ & $H(z)$ & $\sigma_{tot}$ & $H(z)$ & $\sigma_{tot}$ \\
\hline
0.1791 & 75 & 4 & 76 & 7\\
0.1993 & 75 & 5 & 73 & 6\\
0.3519 & 83 & 14 & 88 & 25\\
0.5929 & 104 & 13 & 101 & 13\\
0.6797 & 92 & 8 & 86 & 8\\
0.7812 & 105 & 12 & 106 & 14\\
0.8754 & 125 & 17 & 118 & 14 \\
1.037 & 154 & 20 & 126 & 20\\
\hline
\end{tabular}
\caption{$H(z)$ measurements (in units of [$\mathrm{km\,s^{-1}Mpc^{-1}}$]) and their errors as estimated from the median $D4000_{n}-z$ relations and from
the upper envelope, as defined in the text. 
These values have been estimated with BC03 stellar population synthesis models.}
\label{tab:HzUE}
\end{center}
\end{table}

As discussed in the text, both the selection criteria and the method of analysis have been intended to minimize this effect, 
which is expected to be more important for less massive ETGs, in which a residual of star formation is still ongoing (e.g. see \cite{pozzetti}).
To test the reliability of our measurements, we tackled this issue from two sides.
\begin{enumerate}
\item We re-analyzed our data estimating the Hubble parameter from the upper envelope of the $D4000_{n}-z$ relation. This approach has 
been already applied in the literature \cite{Stern, Simon2005}, and probing only the oldest galaxy population at each redshift it
should provide an estimate of $H(z)$ as close as possible to the unbiased one. 
We have considered the following mass ranges: $log({\cal M/M}_{\odot})>11.25$ for SDSS MGS, $11.65<log({\cal M/M}_{\odot})<11.9$ for LRGs, and 
$log({\cal M/M}_{\odot})>10.6$ for $z>0.4$; these mass ranges are defined so that there is no strong median mass evolution in each subsamples along the spanned redshift range.
For each samples, we considered only the $D4000_{n}$ values above the median $D4000_{n}$ in each redshift bin, and then estimate the median $D4000_{n}-z$ 
relation (with its error). In this way we estimated evaluate the 75 upper percentile of the $D4000_{n}-z$ relation, and we used it as a proxy of the upper-envelope. 
The Hubble parameter has been then estimated on this new sample as described previously, with the only change of the conversion parameter $A$ accordingly to 
the $D4000_{n}$ range spanned. The $H(z)$ estimates obtained in the main analysis and from the upper envelope are compatible within $\sim0.3\sigma$ on average,
and for the majority of the values with differences smaller than 6\%, showing therefore a remarkable agreement. These measurements are shown in Fig. \ref{fig:HzUE} and 
reported in Tab. \ref{tab:HzUE}.

\item We estimated quantitatively this effect, using conservative assumptions based on observational constraints. 
However, since these estimates will be model-dependent and based on constraints coming from other surveys, and since, as just discussed, our data are compatible with 
not being biased by the progenitor-bias, we preferred not to add them to the total error budget,
and we considered pessimistic priors to provide an upper limit estimate to the error due to this effect.

The effect that the progenitor-bias has on the age-redshift relation is illustrated qualitatively in Fig. \ref{fig:Hz_testprogbias}.
Since the low-z ETGs sample are biased towards lower ages with respect to the high-z samples, it basically acts by flattening
the age(z) relation: in this way the ``observed'' $H(z)$ will result bigger than the ``true'' one. This effect can be studied by estimating 
the percentage shift $\Delta_{\rm prog\,bias}$ on the Hubble parameter:
\begin{equation}
H(z)_{\rm observed}=H(z)_{\rm true} (1+\Delta_{\rm prog\,bias})
\end{equation}

\begin{figure}[t!]
\begin{center}
\includegraphics[width=0.47\textwidth]{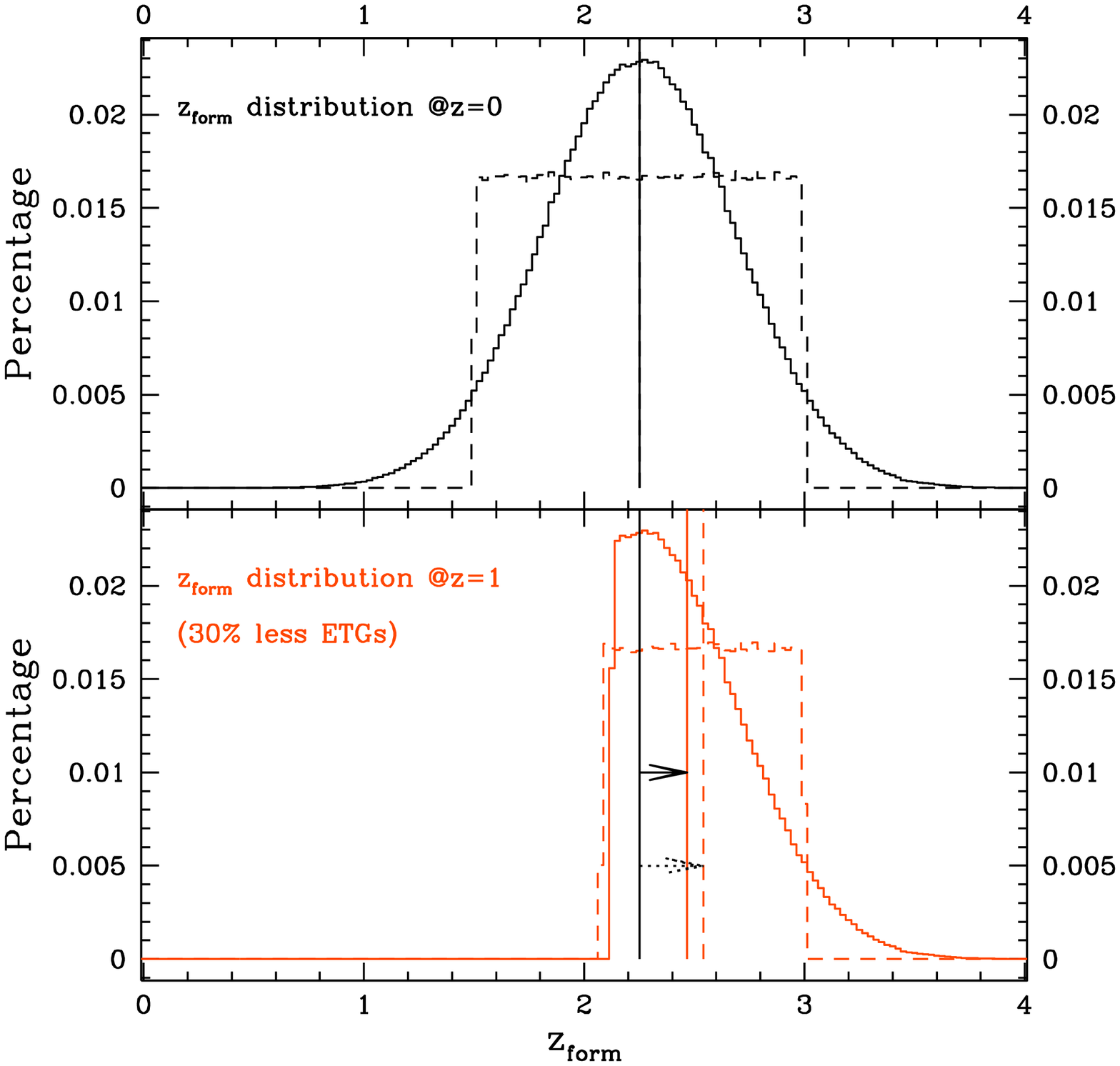}
\includegraphics[width=0.47\textwidth]{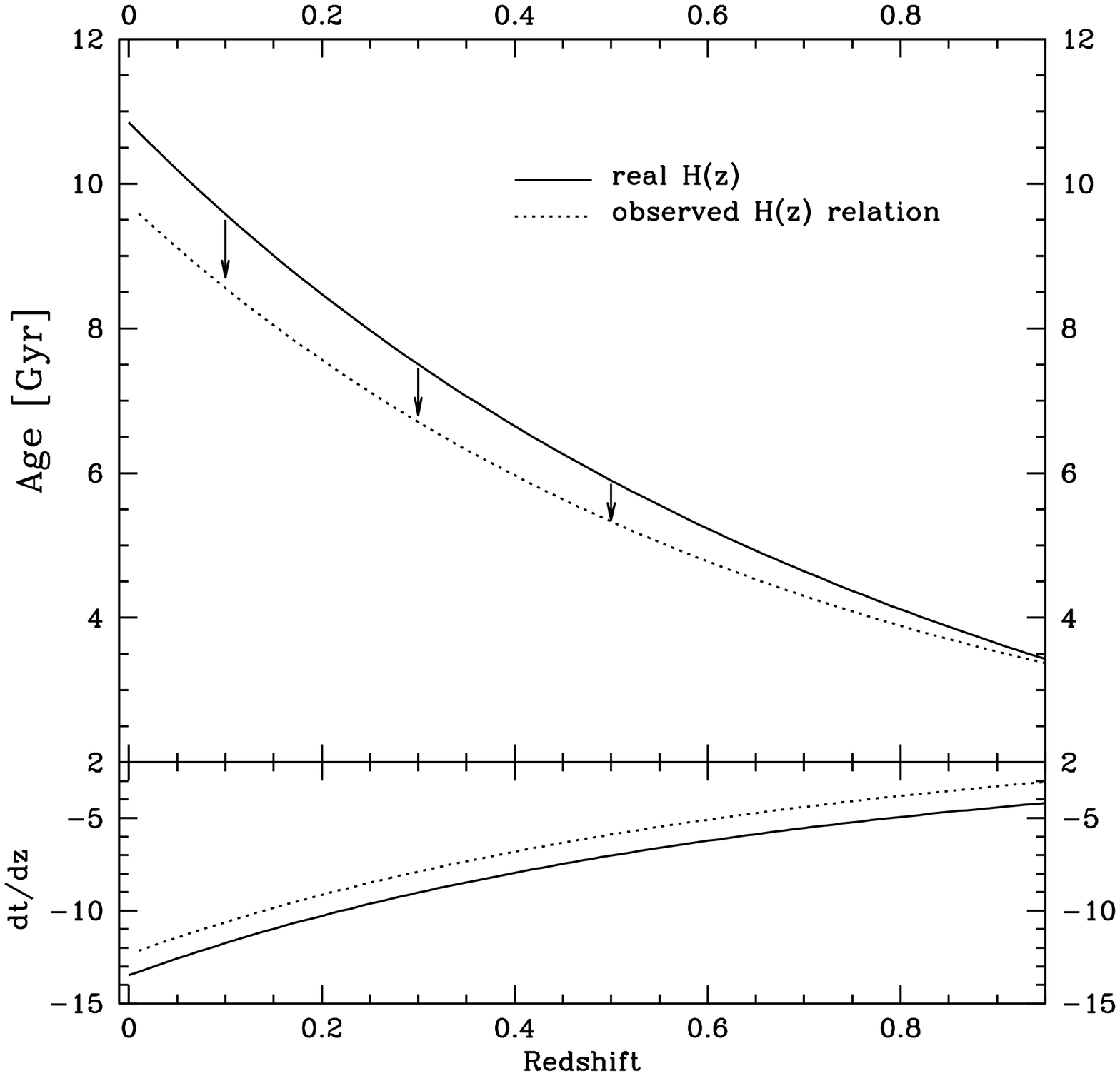}
\caption{Effect of the progenitor-bias on the age-z measurements. Left panel: Hypothetical distribution of redshift of formation
for a sample of ETGs at $z=0$ (upper panel) and at $z=1$ (lower panel). The cases of a gaussian distribution (continuous distribution) and of a flat
distribution (dashed distribution) are considered. Due to the progenitor-bias effect, the left and younger side
of the distribution drops out of the sample at increasing redshift: here an example is shown for which $\sim30$\% of the ETGs are lost
at $z=1$ with respect to $z=0$. The arrows show the variation of the mean redshift of formation (vertical continuous and dashed lines).
Right panel: Biasing the low-z ETGs sample towards lower ages, the observed age(z) relation results flattened with respect to the true one.
In the lower panel is shown the different slope of the two relations.
\label{fig:Hz_testprogbias}}
\end{center}
\end{figure}

The progenitor-bias can be simply considered as an evolution of the mean redshift of formation as a function of redshift
of the ETG samples. In the case that the redshift of formation of the galaxies is homogeneous, i.e. $\Delta z_{\rm form}/\Delta z=0$, then
there will be no progenitor-bias effect; otherwise, if $\Delta z_{\rm form}/\Delta z=B\neq0$, this means that, in a given redshift bin, instead of measuring the real
age difference $dt$ due to the passive evolution, one will measure $dt-\delta$, where $\delta$ will be due to the variation of redshift of formation.
The ``observed'' H(z) relation will be $H(z)_{\rm observed}=-1/(1+z)\frac{dz}{dt-\delta}$, and the percentage shift with respect to the ``true'' Hubble 
parameter will be therefore equal to:
\begin{equation}
\Delta_{\rm prog\,bias}=\frac{H(z)_{\rm observed}}{H(z)_{\rm true}}-1=\frac{1}{1-\delta/dt}-1
\end{equation}

The age of a galaxy $t(z)$ is defined as:
\begin{equation}
t(z)=\int_{z}^{\infty}\frac{dz'}{(1+z')H(z')}
\end{equation}
and hence the age difference $\delta$ between two galaxies formed at redshift $z_{f,1}$ and $z_{f,2}$ is:
\begin{eqnarray}
\delta&=&t(z_{f,1})-t(z_{f,2})=\int_{z_{f,2}}^{z_{f,1}}\frac{dz'}{(1+z')H(z')}\nonumber\\
&\sim&\frac{1}{(1+z_{\rm form})H(z_{\rm form})}\Delta z_{\rm form}\sim\frac{1}{(1+z_{\rm form})H(z_{\rm form})}B\Delta z
\end{eqnarray}
where in the last equation we considered $z_{\rm form}$ as the mean redshift of formation and, as stated before, $\Delta z_{\rm form}/\Delta z=B$. 
Remembering that $dz/dt=(1+z)H(z)$, it is possible to write:
\begin{equation}
\delta/dt\sim\frac{\Delta z_{\rm form}}{\Delta z}\frac{1+z}{1+z_{\rm form}}\frac{H(z)}{H(z_{\rm form})}
\label{eq:Hzprogbiascorrect1}
\end{equation}
and the ``true'' $H(z)$ relation can be recovered from the observed one from:
\begin{equation}
H(z)_{\rm true}=H(z)_{\rm observed}(1-\delta/dt)
\label{eq:Hzprogbiascorrect2}
\end{equation}

To obtain an estimate of $\delta/dt$, we considered a simple model: we assumed that a given distribution of redshifts of formation for ETGs
is observed at $z=0$, and that at increasing redshift the youngest galaxies drop out of the sample due to selection effects, and so the left side of the distribution results
cut. This will produce a variation of the mean redshift of formation as a function of redshift, $\Delta z_{\rm form}/\Delta z$ (see Fig. \ref{fig:Hz_testprogbias}).
Many works have found that, at least up to $z\sim1$, the number density of the most massive ETGs is almost constant, while it
shows a more appreciable decrement with redshift at smaller masses. Ref. \cite{pozzetti}, studying the zCOSMOS survey, reported an evolution of less than 15\% for galaxies
with $log({\cal M/M}_{\odot})>11$, and of $\sim50\%$ for masses $log({\cal M/M}_{\odot})\sim10.5$. Ref. \cite{Scarlata} analyzed ETGs in the COSMOS survey, finding no 
evidence for a decrease in the number density of the most massive ETGs out to $z\sim0.7$; relaxing the assumption about star formation histories and other properties, they 
estimate a maximum decrease in the number density of massive galaxies at that redshift of $\sim30\%$. The recent analysis of Ref. \cite{Brammer2011} found a noticeable evolution 
in number density of quiescent galaxies in the NEWFIRM Medium-Band Survey from $z\sim0$ to $z\sim2$, but this evolution is remarkably smaller up tp $z\sim1$, and also compatible
with no evolution within the errorbars. In order to give an estimate for our two mass subsamples, we assume conservatively an evolution of $\sim30\%$ for ETGs with 
$log({\cal M/M}_{\odot})>11$ from $z\sim0$ to $z\sim1$, and of $\sim50\%$ for ETGs with $11<log({\cal M/M}_{\odot})<10.6$. Not having any prior knowledge about the shape 
of the distribution of the redshifts of formation, we considered the case of a flat distribution (as an extreme case) between $z_{\rm f,min}$ and $z_{\rm f,max}$ and the case of
a gaussian distribution (as suggested e.g. by Ref. \cite{thomas}) with the same mean and dispersion of the flat one. 

We proceeded as follows:
\begin{itemize}
\item we built several distributions of redshifts of formation at $z=0$ considering different combinations of $z_{\rm f,min}$ and $z_{\rm f,max}$, compatible with the observations 
of relatively high redshifts of formation for these massive and passive ETGs, i.e. $(z_{\rm f,min}$,$z_{\rm f,max})=(1,3),(1.5,3),(2,3)$;
\item for each of these models, we considered both the case of a flat and of a gaussian initial distribution;
\item for each of these distributions, we estimated the relation $z_{\rm form}(z)$ between $z=0$ and $z=1$ considering to have at $z=1$ respectively 30\% and 50\% 
less younger ETGs with respect to $z=0$, i.e. for two cases representative of the evolution of the high mass bin ($log({\cal M/M}_{\odot})>11$) and of the low mass bin 
($11<log({\cal M/M}_{\odot})<10.6$);
\item for all the models, we estimated the quantity $\Delta z_{\rm form}/\Delta z$, and we averaged them to obtain a mean $<\Delta z_{\rm form}/\Delta z>$ respectively
for the high- and low-mass bin;
\item we used the estimated $<\Delta z_{\rm form}/\Delta z>$ to evaluate the quantity $\delta/dt$ at the redshifts used in our analysis, considering a flat 
$\Lambda$CDM model ($\Omega_{m}=0.27$, $H_{0}=72\mathrm{\;km\;s^{-1}Mpc^{-1}}$). To estimate this quantity we took into account that the {\it high-mass} sample of
our analysis has at all redshifts $log({\cal M/M}_{\odot})>11$, while the {\it low-mass} sample has $log({\cal M/M}_{\odot})>11$ at $z<0.4$ and 
$10.6<log({\cal M/M}_{\odot})<11$ at $z>0.4$ (see Tab. \ref{tab:D4000med});
\item we estimated a mean error due to the progenitor-bias using Eq. \ref{eq:Hzprogbiascorrect1} and \ref{eq:Hzprogbiascorrect2} for the {\it low-} and {\it high-mass} subsamples 
of ETGs, and averaged them to estimate the mean effect on $H(z)$.
\end{itemize}
The values $\sigma_{\rm prog\,bias}$ obtained are reported in Tab. \ref{tab:errprogbias}. The different model studied have been built to consider the uncertainty in our 
knowledge of the distribution of redshifts of formation, estimating in this way an averaged effect. We notice that the estimated errors are smaller than the statistical errors, on 
average $\sim0.6\sigma_{stat}$. In Tab. \ref{tab:errprogbias} we also show the increase in the total error when also this uncertainty is added in quadrature; it is evident 
that the variation in the total error is really small, adding only $\sim1\%$ on average.

To cross-check the results obtained with the model introduced above, we also studied the theoretical models which predicts the effect of the progenitor-bias from van Dokkum \&
Franx (2001)\footnote{http://www.astro.yale.edu/dokkum/evocalc/} (hereafter vDF, \cite{vanDokkum2001}); we refer to this reference for a detailed discussion about the 
parameters used to set the models. We varied the available parameters building different models, fixing only the amount of ETGs already in place at $z=1$ ($\sim70\%$ for 
the high masses and $\sim50\%$ for the low masses) and the redshift of formation ($2<z_{\rm form}<3$); using these models, we estimated an error $\sigma^{vDF}_{\rm prog\,bias}$ 
as described above for our model. These values are reported in Tab. \ref{tab:errprogbias}. The results obtained with vDF models are compatible with what found with our 
model, showing also in this case that, given our selection criteria, the progenitor-bias add a small contribution to the total errorbar, $\sim2\%$ on average. We also studied the default
``strong evolution'' option provided, finding in this case even smaller errors due to the high value of redshift of formation set.

\begin{table}[h!]
\begin{center}
\begin{tabular}{|c|cc|cc|}
\hline
$z$ & $\sigma_{\rm prog\,bias}$ & \% increase in & $\sigma^{vDF}_{\rm prog\,bias}$ & \% increase in\\
& [$\mathrm{km\,s^{-1}Mpc^{-1}}$] & total error & [$\mathrm{km\,s^{-1}Mpc^{-1}}$] & total error\\
\hline
0.1791 & 2 & 0.8\% & 3 & 1.1\%\\
0.1993 & 2 & 0.7\% & 3 & 0.9\%\\
0.3519 & 3 & 0.3\% & 3 & 0.4\%\\
0.5929 & 6 & 1.3\% & 8 & 2\%\\
0.6797 & 6 & 1.9\% & 7 & 2.8\%\\
0.7812 & 7 & 1.2\% & 9 & 2.2\%\\
0.8754 & 10 & 1.9\% & 12 & 2.8\%\\
1.037 & 10 & 1.5\% & 13 & 2.5\%\\
\hline
\end{tabular}
\caption{Upper limit theoretical estimates of the error due to progenitor-bias (in units of [$\mathrm{km\,s^{-1}Mpc^{-1}}$]) in the various redshifts bin and the percentage increase
in the total error when also this contribution is summed in quadrature. These values have been estimated using both the simple model described here ($\sigma_{\rm prog\,bias}$)
and the models of Ref. \cite{vanDokkum2001} ($\sigma^{\rm vDF}_{\rm prog\,bias}$).}
\label{tab:errprogbias}
\end{center}
\end{table}
\end{enumerate}

As a further check, it is worth recalling that the values of $H(z)$ obtained with this analysis are in very good agreement with the measurements
obtained in literature with other approaches and methods (see Sec. \ref{sec:comparison} and Fig. \ref{fig:HzUE}). 
From these tests we therefore conclude that, given the present-day errorbars, our data are compatible with not being affected by the progenitor-bias.

\subsection {Impact of the adopted Initial Mass Function}
The impact of the IMF on the $D4000_{n}$ is insignificant, and does not affect our analysis: the difference between $D4000_{n}$ values estimated in a
single stellar population model with a Chabrier or a Salpeter IMF are less than 0.3\% for all the metallicities considered in this analysis, and less than 
0.2\% for the solar metallicity (which is the one that better fit our ETGs).

\subsection {The $\alpha-$enhancement}
The effect of the $\alpha-$enhancement is slightly more difficult to be analyzed. Many studies have pointed out that massive ETGs are enhanced in $\alpha$ elements (e.g. see
\cite{alpha1,alpha2,alpha3,alpha4,alpha5,alpha6}) with respect to solar neighborhood. Even if some works \cite{Coelho2007,TMJ} have been developed to take into account 
also this effect in modeling stellar populations, the impact of considering the $\alpha-$enhancement on the $D4000_{n}$ has not yet been modeled and studied into details. 
The models of Ref. \cite{Coelho2007} are available\footnote{http://sites.cruzeirodosulvirtual.com.br/nat/modelos.html} for 6 chemical mixtures, at 3 fixed values of [Fe/H] one model
solar scaled and one $\alpha-$enhanced for which the abundances of all classical $\alpha$ elements are increased by 0.4 dex relative to the scaled-solar model; the corresponding
stellar metallicities are in the range $0.3<Z/Z_{\odot}<3$. For these models, all the different spectral indices are available, included the $D4000_{n}$. We decided to compare the slope
of the $\alpha-$enhanced models with the solar-scaled models to check the impact of this effect; however it has to be underlined that, due to the availability of the models, this 
comparison is different from the study performed in our analysis, since it is done at fixed [Fe/H] instead of at fixed metallicity. We found that, in the range of $D4000_{n}$ relevant 
for our analysis, while the absolute values of $D4000_{n}$ are different, the slopes are not strongly dependent on the $\alpha$-enhancement, with a percentage difference 
between 2\% and 8\%\footnote{These values have been obtained for the case of sub-solar metallicities. For the case of the solar metallicity, the $D4000_{n}$-age relation was
not sampled in the range of $D4000_{n}$ of interest}. This results is confirmed also studying single stellar population models based on MARCS theoretical
libraries \cite{marcs}, using solar metallicity and $[\alpha/Fe]=0,0.4$ (Claudia Maraston, private communication); in this case the difference in $D4000_{n}$ is extremely small,
on average only $\sim0.5\%$, with a percentage variation in the slope of $\sim8\%$. 
As a final remark, it has to be considered also that since in our analysis we consider the most massive ETGs, these galaxies will have quite similar enhancements, so that considering
this effect will not introduce an additional systematic error, but just a shift in the slope. In this way the relative effect will be further mitigated.

%%%%%%%%%%%%%%%%%%%%%%%%%%%%%%%%%%%%%%%%%%%%%%%%%%%%%%%%%%%%%%%%%%%%%%%%%%%%%%%%%%%%%%%%%%%

%%%%%%%%%%%%%%%%%%%%%%%%%%%%%%%%%%%%%%%%%%%%%%%%%%%%%%%%%%%%%%%%%%%%%%%%%%%%%%%%%%%%%%%%%%%

\end{document}